\documentclass[%
 reprint,
frontmatterverbose, 
amsmath,amssymb,
 aps,
 prb,
 showkeys
]{revtex4-2}

\usepackage{graphicx}
\usepackage{dcolumn}
\usepackage{bm}
\usepackage{xspace}
\usepackage[nolist, nohyperlinks]{acronym}
\usepackage[colorlinks]{hyperref}

\usepackage{textcomp}
\usepackage{gensymb}
\usepackage[dvipsnames]{xcolor}
\usepackage{enumerate}

\usepackage[colorinlistoftodos, textsize=tiny,backgroundcolor=Goldenrod]{todonotes}

\newcommand{\Fe}[1]{[#1]$_{\mathrm{Fe}}$}

\newcommand{\Vu}[1]{\ensuremath{\hat{#1}}}
\newcommand{\VV}[1]{\ensuremath{\bm{#1}}}
\newcommand{\vni}{\ensuremath{\Vu{n}_i}}
\newcommand{\vnj}{\ensuremath{\Vu{n}_j}}
\newcommand{\vrij}{\ensuremath{\Vu{r}_{ij}}}
\newcommand{\Beff}{\ensuremath{\frac{1}{\mu_i}\frac{\partial\mathcal{H}}{\partial\vni}}}
\newcommand{\DKperp}{\ensuremath{\Delta K_{1,\perp}}}
\newcommand{\Koneperp}{\ensuremath{K_{1,\perp}}}
\newcommand{\fres}{\ensuremath{f_\mathrm{res}}}
\newcommand{\Kshapeoop}{\ensuremath{K_{\textrm{shape},\perp}}}
\newcommand{\Kpshapeoop}{\ensuremath{K'_{\textrm{shape},\perp}}}
\newcommand{\Kpshapeip}{\ensuremath{K'_{\mathrm{shape},\parallel}}}
\newcommand{\TClayer}{\ensuremath{T_{\mathrm{C,\,layer}}}}
\newcommand{\TCtwodots}{\ensuremath{T_{\mathrm{C,2\,dots}}}}
\newcommand{\TSRT}{\ensuremath{T_{\mathrm{SRT}}}}
\newcommand{\Mslayer}{\ensuremath{M_{\mathrm{s,\,layer}}}}
\newcommand{\mlayer}{\ensuremath{m_{\mathrm{layer}}}}
\newcommand{\mlayerperp}{\ensuremath{m_{\mathrm{layer,\perp}}}}

\begin{document}

\preprint{APS/123-QED}

\title{Multiscale approach for modeling magnetization properties of inhomogeneous ultrathin magnetic layers}
                                   
\author{Julien Mordret}            
\author{Jean-Christophe Le Breton} 
\author{Gabriel Delhaye}           
\author{Bruno Lépine}              
\author{Philippe Schieffer}        
\author{Sylvain Tricot}            

\affiliation{
    Univ Rennes, CNRS, IPR (Institut de Physique de Rennes) - UMR 6251, 
    F-35000 Rennes, France
}

\email{sylvain.tricot@univ-rennes.fr}

\date{\today}

\begin{abstract}
We report on spin atomistic calculations used to model static and dynamic
magnetic properties of inhomogeneous ultrathin iron films.  Active magnetic
layers in next-generation spintronic devices are becoming so thin that they
exhibit some variable degree of roughness at the low-scale making them
magnetically inhomogeneous.  We propose a multiscale approach to progressively
shift from a rough atomic-scale system to an ensemble of macrospins.  By
studying nanoscale islands of atoms in contact with each other, we demonstrate
that ultrathin rough layers can be described by a set of macrospins coupled by
a Heisenberg-like exchange interaction driven by the existence and shape of
nanoconstrictions linking the islands.  We show that the magnetization dynamics
at 0~K is strongly impacted by this surface morphology since the resonant
frequency of a typical ultrathin iron layer can drop by up to an order of
magnitude due to inhomogeneities.  Additionally, we used Monte Carlo
simulations to determine the ferromagnetic-paramagnetic and spin reorientation
transition temperatures for various morphology parameters and we show how
nanoconstrictions and shapes of the atomic clusters can modify these transition
temperatures.  Our results demonstrate the possibility to account for the
morphology of ultrathin structures with significant roughness. We believe that
our approach makes it possible to model complete devices as close as possible
to experimental reality.
\end{abstract}






                              
\maketitle


\begin{acronym}[base]
\acro{MTJ}[MTJ]{magnetic tunnel junction}
\acro{PMA}[PMA]{perpendicular magnetic anisotropy}
\acro{STT}[STT]{spin transfer torque}
\acro{MRAM}[MRAM]{magnetic random access memory}
\acrodefplural{MRAM}[MRAMs]{magnetic random access memories}
\acro{SOT}[SOT]{spin-orbit torque}
\acro{VCMA}[VCMA]{voltage controlled magnetic anisotropy}
\acro{FFT}[FFT]{fast Fourier transform}
\acro{FMR}[FMR]{ferromagnetic resonance}
\acro{FDM}[FDM]{finite differences method}
\acro{FEM}[FEM]{finite elements method}
\acro{FMM}[FMM]{fast multipole method}
\acro{LLG}[LLG]{Landau-Lifshitz-Gilbert}
\acro{LLB}[LLB]{Landau-Lifshitz-Bloch}
\acro{STM}[STM]{scanning tunneling microscope}
\acro{IEA}[IEA]{International Emerging Actions}
\acro{MC}[MC]{Monte-Carlo}
\acro{MCS}[MCS]{Monte-Carlo steps}
\acro{CMC}[CMC]{constrained Monte-Carlo}
\acro{DDI}[DDI]{dipole-dipole interactions}
\acro{MOKE}[MOKE]{magneto-optical Kerr effect}
\acro{SRT}[SRT]{spin reorientation transition}
\acrodefplural{SRT}[SRTs]{spin reorientation tansitions}
\end{acronym}

\begin{acronym}[composed]
    \acro{STT-MRAM}[STT-MRAM]{spin transfer torque \acl{MRAM}}
    \acrodefplural{STT-MRAM}[STT-MRAMs]{spin transfer torque \aclp{MRAM}}
    \acro{SOT-MRAM}[SOT-MRAM]{spin-orbit torque \acl{MRAM}}
    \acrodefplural{SOT-MRAM}[SOT-MRAMs]{spin-orbit torque \aclp{MRAM}}
    \acro{NUFFT}[NUFFT]{non-uniform \ac{FFT}}
    \acro{ACMOKE}[ac-MOKE]{alternative-current \acl{MOKE}}
\end{acronym}

\section{\label{sec:introduction}Introduction}

Ultrathin magnetic layers just a few nanometers thick are an essential building
block of groundbreaking technologies for data storage, non-volatile magnetic
memory like \ac{STT-MRAM} \cite{slonczewski1996, stiles2002, jalil2007,
santos2020, huang2021} or \acp{SOT-MRAM} \cite{miron2011, emori2016},
magnonics-based devices for data processing and transport \cite{chumak2017}, or
other innovative devices \cite{duan2008, rana2019}.  Such low thicknesses give
rise to new properties essentially driven by surface and interface phenomena.
One example is the appearance of \ac{PMA}, widely used in oxyde-based \acp{MTJ}
\cite{monso2002, rodmacq2003, rodmacq2009, lee2011, yang2011, oh2014, peng2015,
dieny2017, beik2019, kowacz2021}.  The effect arises due to some hybridization
between out-of-plane $3d$ wave function of the ferromagnetic layer with the O
$2p$ orbitals of the oxide leading to a considerable enhancement of the
\ac{PMA} \cite{dieny2017}. The effect can even be further enhanced by interface
hybridization with a heavy metal capping layer \cite{peng2015} or by exchange
bias with an antiferromagnetic metal capping layer \cite{wang2011, wang2013}.

However, the growth of ultrathin metal films onto metal, semiconductor
substrates or oxides often leads to rough surfaces \cite{giergiel1995,
shen1995, schaller1999, martinez2005, lin2009, torelli2009, chien2012,
herve2013, ye2014} due to surface energies considerations and to the growth
process occurring close to the thermal equilibrium.  And no matter how much
care is taken to achieve growth, a roughness of just 2 atomic steps, which
would be perfect for a 50~nm film, becomes critical for a film of only few
atomic planes.  This disturbed surface morphology can have a serious impact on
static or dynamic magnetic properties of these ultrathin films.  Some studies
showed that, for thin layers of FeNi alloys onto a Si(100) substrate, the
surface roughness results in a lower resonant frequency than would be expected
for a smooth film \cite{azevedo2000}. Other groups also showed that the more
the surface is rough, the lower the Curie temperature \cite{schneider1990,
ohresser1999, spangenberg2005, wenchin2006, bauer1997} or that \ac{PMA}
properties are modified when the surface morphology changes \cite{schaller1999,
kim2001, enders2003}.  Many experimental observations have also shown that
roughness actually tends to favor in-plane magnetization configurations
\cite{schaller1999, kim2001, enders2003}.

\begin{figure}[htbp]
\includegraphics[width=86mm]{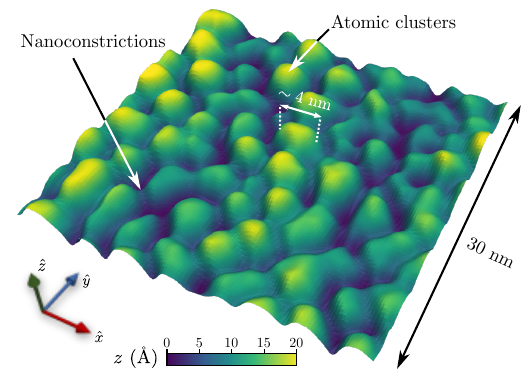}
\caption{\label{fig:roughness} 
    [Color online] 3D representation of a 1.4~nm thick Fe deposited at room
    temperature on SrTiO$_3$(001).  The surface is composed of small atomic
    clusters $\sim 4-5$~nm in diameter densely interconnected by
    nanoconstrictions (data extracted from \ac{STM} images \cite{chien2012}).
}
\end{figure}

Figure~\ref{fig:roughness} is a three dimensional representation of
$\sim$10~monolayers of Fe deposited on a SrTiO$_3$(001) substrate obtained from
a \ac{STM} study (data extracted from \cite{chien2012}).  The actual surface
morphology is such that the metallic film can be viewed as a dense packing of
atomic clusters of atoms, roughly 4~nm in diameter, self-arranged in a dense
lattice close to a triangular 2D-lattice.  As can be seen in
figure~\ref{fig:roughness}, clusters are connected to each other by only few
atoms forming an atomic constriction.  A survey of the literature reveals that
many ferromagnetic ultrathin films exhibit morphologies very similar to the one
presented in figure~\ref{fig:roughness} \cite{giergiel1995, shen1995,
schaller1999, martinez2005, lin2009, torelli2009, chien2012, herve2013,
ye2014}.  We will indifferently use the words \textit{islands} or \textit{dots}
and \textit{nanoconstriction} or \textit{atomic bridge} to refer to the
clusters and atomic constrictions respectively.  

In this article, we propose to study, through numerical simulations, how this
so peculiar surface morphology could affect static and dynamical magnetic
properties of an ultrathin Fe film.  On the one hand, neither \ac{FDM} nor
\ac{FEM} can be used for simulating these systems given the pronounced
inhomogeneity due to this surface morphology \cite{bruno1988, zhao1999}. On the
other hand, classical spin atomistic methods are too resource-intensive to
describe these systems at a large scale.
Therefore, we propose here a multiscale approach based on classical atomistic
spin simulations to model the dynamic at 0~K and static magnetic properties of these
inhomogeneous films. 
Such a multiscale approach combined with \ac{LLB} model was used to simulate
magnetization dynamics in FePt and FeNi alloys \cite{kazantseva2008, hinzke2015}.
We have chosen to focus on the magnetic properties of a 
$\sim$1~nm thick Fe(001) film having \ac{PMA} at 0~K. Inspired by 
figure~\ref{fig:roughness}, we
have modeled these inhomogeneous ultrathin films as a set of macrospins
exchange-coupled through nanoconstrictions. This allows us to treat
inhomogeneous systems on a larger scale, taking into account all interactions,
in particular the inter-island \ac{DDI}, which is usually
numerically very demanding.  We specifically studied the effect of changing the
film morphology on the resonant frequency at 0~K, as well as on the
ferromagnetic-paramagnetic transition and spin-reorientation temperatures.

A brief description of the equations and the methods used in atomic-scale spin
simulations are first recalled in section~\ref{sec:Method}. The main matter
follows in section~\ref{sec:results} split in three parts. We first study the
magnetic configuration of two islands connected together by nanoconstrictions
at the atomic scale in section~\ref{sec:macrospins}. In particular, we show
that these highly inhomogeneous ultrathin films can be described in a macrospin
approach, where clusters are treated as macrospins interacting via a Heisenberg
interaction at $T=0$~K whose exchange constant is directly related to the shape
of the atomic link between the islands. This result makes it possible to apply
a scaling law to the atomistic description and simulate highly inhomogeneous
systems at the atomic scale over lengths exceeding 1000~nm.  This approach is
used in section~\ref{sec:mag_dynamics} to model the dynamical magnetic
properties of an ultrathin ferromagnetic iron layer whose inhomogeneity
properties are close to a realistic scenario. We focus on the resonant
frequency of such a system and highlight the extent to which this inherent
low-scale roughness can impact the magnetization dynamics. In
section~\ref{sec:Temperature}, we show how morphologies can affect temperatures
of both ferromagnetic to paramagnetic and spin reorientation transitions. In
particular, we disentangle the impact of nanoconstrictions and cluster shapes
in these processes using atomistic spin \ac{MC} simulations and a multiscale
approach.  Finally, in section~\ref{sec:discussion} we discuss the limitations
and improvements of our approach, before concluding in
section~\ref{sec:conclusions}.

\section{\label{sec:Method}Computational details}

Calculations were done within the atomistic spin model where an ensemble of
classical spins located on lattice sites $i$ are interacting through an
extended Heisenberg hamiltonian of the form~\cite{muller_spirit_2019}:
\begin{equation}
    \label{eqn:hamiltonian}
    \begin{split}
        \mathcal{H} = &-\sum_i \mu_i \mu_0\VV{H_0}\cdot\vni -\sum_{<ij>} J_{ij} \vni \cdot \vnj \\
                    &-\sum_i \sum_j k_i\left(\Vu{k}_{j, \textrm{at}}\cdot\vni\right)^2  -\sum_i \sum_{j=1}^{3} k_{4,\textrm{at}}\left(\Vu{u}_j\cdot\vni\right)^4 \\
                    &-\frac{1}{2} \frac{\mu_0}{4\pi} \sum_{\substack{i,j \\ i \neq j}} \mu_i \mu_j \frac{3(\vni \cdot \Vu{r}_{ij})(\vnj \cdot \Vu{r}_{ij}) - \vni \cdot \vnj}{r_{ij}^3},
    \end{split}
\end{equation}
where $\Vu{n}_{i,j}$ are unit vectors of the onsite atomic magnetic moments $i$
and $j$, of norm $\mu_i$ for atom $i$. The above equation is composed of five
competing energy terms where the first one is the Zeeman energy with \VV{H_0}
being the externally applied magnetic field, the second one is the exchange
energy defined by the Heisenberg exchange constant $J_{ij}$ that is to be
considered for every $<ij>$ unique pair of neighboring interacting spins $i$
and $j$.  Third and fourth terms are the magnetic anisotropy energies for spin
$i$ with anisotropy constants $k_{j,\textrm{at}}$ for every uniaxial anisotropy
$j$ of easy axis $\Vu{k}_{j,\textrm{at}}$ and $k_{4,\textrm{at}}$ is the
anisotropy constant for the cubic magnetocrystalline anisotropy energy defined
along the basis unit vectors $\Vu{u}_j$ (corresponding to the [100], [010] and
[001] directions of the Fe lattice in the following, respectively along
$\Vu{x}$, $\Vu{y}$ and $\Vu{z}$ directions of the basis).  In the macrospin
approach, expression~\ref{eqn:hamiltonian} remains the form used for
calculations such that several uniaxial anisotropy terms can co-exist, as they
allow us to account for magnetocrystalline anisotropy (\ac{PMA}), as well as
island shape anisotropy.  The last term is the energy due to the demagnetizing
field, modeled as dipolar interactions between spins $i$ and $j$ where their
joining vector is of unit direction $\vrij$ and norm $r_{ij}$.  $\mu_0$ is the
vacuum permeability.  The time evolution of the magnetization is computed by
solving the \ac{LLG} equation~\cite{landau1935, gilbert2004}, which reads
\begin{equation}\label{eqn:llg}
\begin{split}
    \frac{\partial\vni}{\partial t} = &\frac{\gamma}{(1+\alpha^2)} \vni \times \Beff \\
                                      & + \frac{\gamma\alpha}{(1+\alpha^2)} \vni \times \left( \vni \times \Beff \right).
\end{split}
\end{equation}
Equation~\ref{eqn:llg} includes terms for precession and damping of the
magnetic moment where $\gamma$ is the electron gyromagnetic ratio and $\alpha$
is a damping parameter.  We used the \textsc{Spirit}
framework~\cite{muller_spirit_2019} for computing the spin dynamics with a
damping parameter $\alpha$ = 0.7--1.0 for static solutions and $\alpha = 0.008$
for time-resolved calculations.  First neighbors were considered in the
exchange interaction with $J_1=27$~meV \cite{kvashnin2016, cardias2017}.  When
included the exchange constant of the second-neighbor contribution is
$J_2=19$~meV \cite{cardias2017}.

\Ac{FFT} method was used to compute the dipolar field. The timestep was no more
than 0.1~ps and the convergence criterion is defined as: $\max || d\Vu{n}_i ||
\leq \epsilon \; \forall i$, where $\Vu{n}_i$ is the unit magnetization vector
of spin (or macrospin) $i$ and $\epsilon$ is a threshold value.  This quantify
is unitless and proportional to the asymptotic magnetic torque. In our
calculations, $\epsilon \leq 4\times 10^{-3}$.

\section{\label{sec:results}Results}

\subsection{\label{sec:macrospins}From atomistic spins to macrospins}

\begin{figure}[htbp]
\includegraphics[width=86mm]{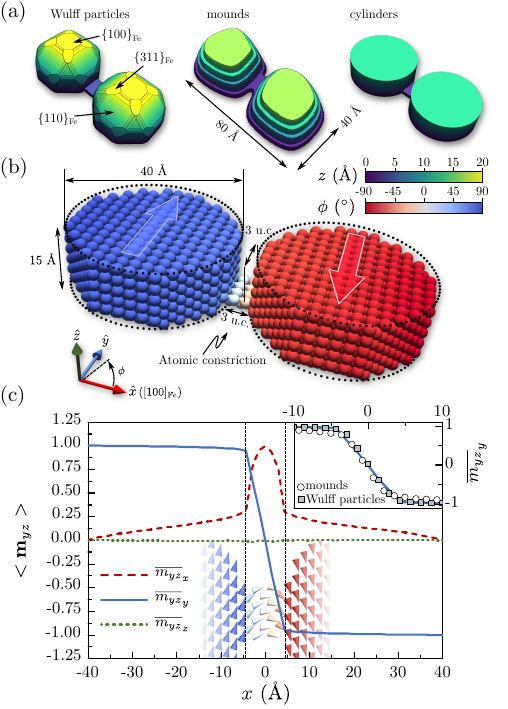}
\caption{\label{fig:fig1} 
    [Color online]
    a) The three different shapes used to model two clusters connected by a
    narrow atomic constriction. b) 3D view of the two Fe cylinders system used
    for the magnetic atomistic calculations. The color map for atoms corresponds
    to the magnetization angle $\phi$ in the $xy$-plane where $\phi=0$ is
    the \Vu{x} direction. c) Magnetization of the two cylinders system shown
    above averaged over the $yz$-plane projected along \Vu{x}, \Vu{y} and
    \Vu{z} directions.
}
\end{figure}

To understand the magnetic properties of thin films or micro/submicro
structures with this irregular surface morphology, we are first interested in
the magnetic configuration of two Fe clusters connected by a narrow atomic
constriction at $T=0$~K. Each cluster is composed of approximately 1500~atoms
and, although the exact shape of a cluster itself is not the focus of this
paper, it stems from growth conditions of which two opposing kinds can be
distinguished leading to the formation of 3D islands.  \textit{i)} when the
growth is purely driven by the thermodynamics, the shape of the island is
controlled by surface and interface energies of the iron crystal. Growth rates
of specific atomic plane directions are different resulting in a geometric
construction in which some facets are favored (Wulff construction).  We
reproduced this shape in figure~\ref{fig:fig1}a (left image) for iron onto a
silicon substrate \cite{rahm2020}.  

Surface energies are taken from \cite{tran2016}. The particle surface is mostly
composed of \{100\} and \{110\} family of planes leading to a shape resembling
a half-sphere 20~\AA\ in radius.  \textit{ii)} The cluster shape may otherwise
be obtained by considering that the growth occurs far from equilibrium.
Mechanisms where the growth is driven by kinetic instabilities imply the
presence of an additional activation energy to downhill diffusion for adatoms
lying at the edge of a terrace. This energy barrier known as the
Ehrlich–Schwoebel (ES) barrier promotes a 3D growth where islands are in the
shape of mounds or pyramids \cite{stroscio1995, koch1999}
(figure~\ref{fig:fig1}a middle image).  In that case, the atomic constriction
joining the Fe dots is similar to a discontinuous wetting layer.

\subsubsection{Magnetic configuration of two Fe islands connected by a nanoconstriction} 
\label{sec:two_islands_magnetization} 

The true shape of the clusters is neither quite a Wulff particle nor quite a
mound and we propose to model it by a cylinder 40~\AA\ in diameter and
13.3~\AA\ in thickness (right image of figure~\ref{fig:fig1}a). For all three
shapes, atomic constrictions contain between 17 to 20 atoms.

Figure~\ref{fig:fig1}b shows the magnetization reversal in two Fe cylinders
connected by a nanoconstriction. Each cluster is composed of 1485~Fe atoms
arranged in the \textit{bcc} structure of iron (lattice parameter = 2.87~\AA).
Magnetic moments of atoms lying in both ends of the system and on length along
\Vu{x} of 2~\AA\ are pinned. We hereby force the magnetization to rotate
180\degree\ on the distance corresponding to two dots.  This calculation is the
result of iteratively solving equation~\ref{eqn:llg} with $2\times 10^6$
iterations and with a timestep of 0.1~ps.  First neighbors exchange interaction
($J_1=27$~meV \cite{kvashnin2016, cardias2017}), \ac{DDI} and cubic
magnetocrystalline anisotropy ($k_{4,\textrm{at}} = 3.5~\mu$eV/atom) are taken
into account.  
For these systems, we have carefully checked that the magnetic configuration at
0~K is the same whether or not second neighbors are considered. For the sake of
simplicity, we will restrict ourselves to first neighbors in this first part at
0~K. Second neighbors will be included in calculations at $T \neq 0$~K.

The remarkable result of figure~\ref{fig:fig1}b is that all the magnetization
rotation seems to be in the atomic constriction bridging the two dots. We can
also see that, for a given $x$ coordinate, magnetic moments in the $yz$-plane
share the same orientation.  It is then possible to restrict the analysis to
the $x$ dimension alone by looking at the reduced magnetization averaged over
the $yz$-plane ($<\VV{m}_{yz}>$) as a function of $x$ as it is presented in
figure~\ref{fig:fig1}c where each component of $<\VV{m}_{yz}>$ (namely
$\overline{m_{yz}}_x$, $\overline{m_{yz}}_y$ and $\overline{m_{yz}}_z$) is
plotted as a function of $x$, with $x=0$ being the center of the atomic bridge.
It is clear that, $\overline{m_{yz}}_y$ jumps from +1 to -1 in the atomic
constriction even though its length is roughly 3$\times$ shorter than the Fe
exchange length. The $z$-component remains almost null on the whole $x$ range
indicating that the rotation lies within the substrate's plane (Néel like wall)
due to the demagnetizing field which contribute to reduce the total energy as
compared to a Bloch like wall type. But the overall relative energy difference
between these two wall types remains extremely small (of the order of 0.002\%),
since it is associated with the few atoms in the constriction. The
$\overline{m_{yz}}_y$ component for the cylinder is redrawn in the inset of
figure~\ref{fig:fig1}c (solid line) along with the results for Wulff particles
and mounds (square and circle symbols respectively). Values for all three
shapes are close enough so as to justify \textit{a posteriori} the cylinder
shape to model metallic islands in ultrathin films.

The previous two-dots system may however appears too constrained to accurately
represent a thin film composed of closely spaced clusters whose magnetization
is free to rotate. Whether the rotation of the magnetization is still confined
in atomic constrictions when none of the spins are pinned inside a cluster may
be addressed by simply looking at an array of three dots arranged along the
[100]$_{\rm{Fe}}$ direction. Magnetic configurations for such a system clearly
demonstrate that the magnetization rotation is still mainly absorbed by the
atomic bridges (see appendix~\ref{apx:three_dots}).

\subsubsection{Effective exchange between two Fe islands connected by a nanoconstriction} 
\label{sec:two_islands_exchange} 

\begin{figure}[htbp]
\includegraphics[width=86mm]{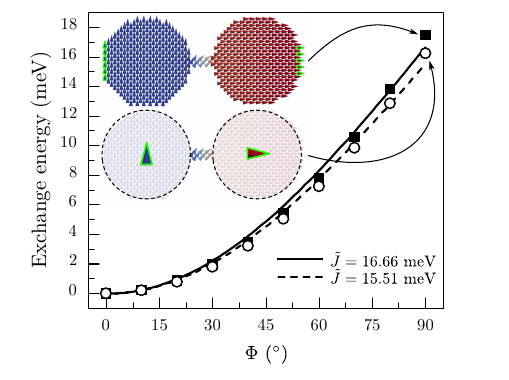}
\caption{\label{fig:fig3} 
    [Color online]
    Total exchange energy with respect to the angle $\Phi$ constraining
    boundary moments (highlighted by a green outline in the inset).
    Calculation values are given with scatter symbols for the fully atomistic
    system (black squares, top image of the inset) and for a hybrid system with
    two macrospins linked by the same atomic bridge than the former system
    (open circles, bottom image of the inset).  The solid and dashed lines
    correspond to a fit with a Heisenberg exchange model with effective
    exchange constant $\tilde{J}$.
}
\end{figure}

The above results suggest that one can approximate the magnetization of a
cluster by a single magnetic moment, the magnitude of which corresponding to
the sum of all Bohr magnetons contained in a dot. In figure~\ref{fig:fig3}, we
computed the exchange energy of two Fe cylinders bridged by a constriction as a
function of the angle $\Phi$ between pinned spins at both ends. 

A comparison between an atomically resolved system (full-square symbols) and a
system where clusters are replaced by a single magnetic moment (macrospin)
while the bridge is still composed of atoms (open-circle symbols) is proposed.
\ac{DDI} were not taken into account in these calculations since they are
negligible in such constrained systems.  Energy values for both systems are
really close and only relatively differ by 7\% at 90\degree.  This confirms
that reducing islands magnetization to a single magnetic moment do not
significantly alter the energy balance.  

The fact that the bridge length is aligned with the [100] direction of the iron
lattice provides a simple way to analytically evaluate the total exchange
energy of such system. As we showed in figure~\ref{fig:fig1}, spins are almost
parallel to each other in any $yz$-plane. The exchange energy associated by any
pair in those planes won't vary when the orientation of the magnetization of
the right dot is changed with respect to the left dot, but exchange interaction
between planes in the $x$ direction will. Let's call $\Delta E_{\textrm{ex}}$,
the exchange energy difference between the case where boundary spins define an
angle $\Phi$ and the case where this angle is zero. If we define by
$n_{\rm{planes}}$ the number of \{100\} planes in a constriction to which we
add both end spins, the number of \textit{inter}-planes contributing to the
total exchange energy will be $n_{\rm{planes}}-1$.  It follows that $\Delta
E_\textrm{ex}$ may simply be written:
\begin{subequations}
    \label{eqn:ExBridge}
    \begin{align}
        \Delta E_\textrm{ex} & = -J_1 p(n_{\rm{planes}}-1)\left[\cos\left(\frac{\Phi}{n_{\rm{planes}}-1}\right) - 1\right] \\
                             & \approx  J_1 \frac{p}{(n_{\rm{planes}}-1)}\frac{\Phi^2}{2}.
    \end{align}
\end{subequations}
In this expression, only first nearest neighbors are considered, $J_1$ is the
Heisenberg exchange interaction constant and $p$ is the averaged number of
pairs per \{100\} plane involved in the exchange interaction. For \textit{bcc}
iron of lattice parameter $a_0$ in [100] direction, $p = 4\times S_b/a_0^2$.
The equation~\ref{eqn:ExBridge} only holds if the cross section of the atomic
constriction is constant over $x$.  Using a simple Taylor expansion to the
second order yields to the approximate expression equation~\ref{eqn:ExBridge}b.
This parabolic form of the exchange energy can then be identified to a
quadratic Taylor approximation of the Heisenberg exchange interaction where the
\textit{effective} Heisenberg exchange constant $\tilde{J}$ of the whole system
of two connected dots is:
\begin{equation}\label{eqn:Jtilde}
    \tilde{J} = J_1\frac{p}{n_{\rm{planes}}-1}
              = J_1\frac{2wh_b}{la_0}
\end{equation}
for bridge geometry parameters $w$, $h_b$, $l$.  Both datasets of
figure~\ref{fig:fig3} were fitted by a Heisenberg exchange interaction between
two magnetic moments of the form $\tilde{J}\left(1 - \cos(\Phi)\right)$.
Adjusted curves along with exchange constants extracted from the model are
reported on figure~\ref{fig:fig3}.  Values are between 15.5 and 16.7~meV for
this bridge geometry. The value from equation~\ref{eqn:Jtilde} is 20.3~meV
which is fairly close.  The discrepancy comes from an overestimation of the
number of pairs connected to each macrospin in the total number of exchange
pairs to take into account ($p(n_{\rm{planes}}-1)$ term in
equation~\ref{eqn:ExBridge}a).

These results demonstrate that the magnetic configuration of two clusters
bridged by an atomic constriction can be approximated by two macrospins
connected by a Heisenberg-like exchange interaction whose exchange constant
$\tilde{J}$ is proportional to the atomic constriction cross section area and
inversely proportional to the length of this constriction.  We also show that a
Heisenberg-like interaction between macrospins is valid as long as angles
between macrospin moments remain $\lesssim 90\degree$ at 0~K.

The results of this first part show that the magnetic configuration of an
ultrathin ferromagnetic film can be described using a Hamiltonian of the same
form as for the atomistic description in equation~\ref{eqn:hamiltonian}, but
where the constants of atomic origin are replaced by effective values derived
from the rough nature of the film.  Atomic magnetocrystalline anisotropy
constants and magnetic moments are multiplied by the number of spins contained
in an island, and the atomic exchange constant $J_1$ is replaced by an
effective constant $\tilde{J}$ containing the geometry of the atomic
constriction. This approach acts as a scaling factor of $\sim$1000, making it
possible to describe a highly inhomogeneous nanoscale system over significant
distances.

\subsection{Magnetization dynamics} \label{sec:mag_dynamics} 

In this section, we propose to use our model to study the \ac{FMR} properties
of ultrathin magnetic films. Our aim is to quantitatively discuss the impact of
an irregular surface morphology with inhomogeneous magnetic properties to the
dynamical magnetic response of systems having \ac{PMA}.

Following the previous discussion, iron clusters are modeled as macrospins with
a magnetization easy axis along \Fe{001}.  The base unit cell of the system is
composed of $26 \times 15 = 780$ macrospins of saturation magnetization $M_s =
1720 \times 10^3$~A/m located onto the nodes of a triangular lattice of
parameter $a=4.5$~nm.  Each island is interacting with its 6 first nearest
neighbors through an Heisenberg-like exchange interaction of constant
$\tilde{J}$ which we varied between 5~meV and 90~meV.  The effect of the
demagnetizing field is taken into account through \ac{DDI} between macrospins
and by assuming that clusters can be treated as cylinders of height
$h_c$=1.2~nm and diameter $D_c$=4.5~nm (volume $V$) leading the
self-demagnetizing energy of an island to be modeled as a shape anisotropy (of
constant $K_{\textrm{shape},\perp}$ in the following equation~\ref{eq:EA1}).
The three elements of the demagnetizing field tensor are thus
$N_{xx}=N_{yy}=0.19$ and $N_{zz}=0.62$ \cite{beleggia2004}.  Periodic boundary
conditions are applied along \Vu{x} and \Vu{y}.  The inset of
figure~\ref{fig:fig4}a shows a sketch of the system.  

\subsubsection{Effect of the nanoconstrictions on the resonant frequency}
\label{sec:fres_exchange}

For now, we consider that clusters are only having a single
uniaxial magnetic anisotropy axis along \Fe{001}. The magnetic anisotropy
free energy density for each island is 
\begin{equation}\label{eq:EA1}
    W_A = -(K_{1,\perp} + K_{\textrm{shape},\perp}) \cos^2 \theta,
\end{equation}
with $\theta$ being the polar angle of the spherical coordinate system and 
$K_{1,\perp}$ being the anisotropy constant including all volume, surfaces
and interfaces contributions from magnetoelastic and magnetocrystalline origins
($K_{1,\perp} > 0$ which favors an out-of-plane magnetization).
$K_{\textrm{shape},\perp}$ is the anisotropy constant related to the shape of a
cluster.  $K_{\textrm{shape},\perp} = \frac{1}{2}\mu_0 M_s^2(3N_{yy}-1) < 0$
which rather promotes an in-plane magnetization. For the simulations,
$K_{1,\perp} = 1.42 \times 10^6$~J/m$^3$ \cite{koo2013} ($K_{1,\perp}V$=170~meV
per macrospin).

\begin{figure}[htbp]
\includegraphics[width=86mm]{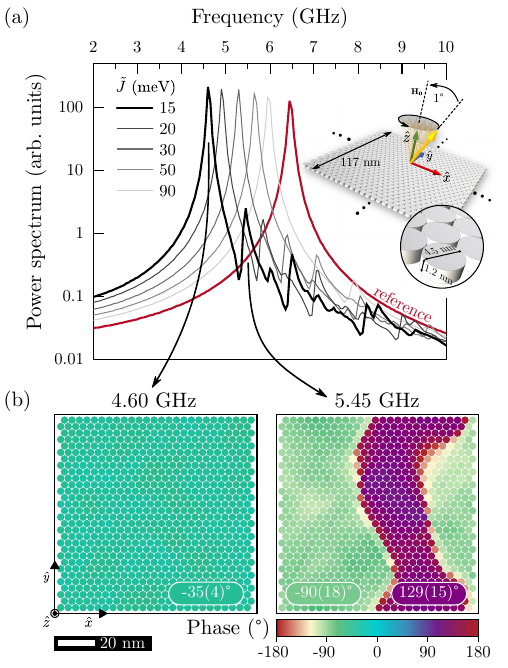}
\caption{\label{fig:fig4}
    [Color online]
    a) Linear-log plot of the spatially averaged power spectra versus frequency
    for different values of $\tilde{J}$ for the inhomogeneous system ($\Delta
    K_{1,\perp}$=30~meV) [curves in grey tones] along with the spectra of the
    homogeneous system with $\tilde{J}$=90~meV [red curve]. Inset: sketch of the 780
    macrospins system arranged onto a periodic triangular lattice. Magnetic
    moments [yellow arrow] are precessing close to the \Vu{z} direction during
    the ringdown run.
    b) $xy$-plane maps of the phase values for the first two resonance
    peaks for $\tilde{J}$=15~meV. Averaged phase value (and standard deviations in
    parenthesis) are labelled on different zones of the map for the uniform
    mode [left map] and a non-uniform mode [right map].
}
\end{figure}

\Ac{FMR} spectra shown in figure~\ref{fig:fig4}a were computed with the ringdown
technique \cite{mcmichael2005, baker2017}. 

The system is first initially relaxed with $\mu_0 H_0 = 0.1$~T slightly
misaligned by 1\degree\ from the \Fe{001} direction.  For this step, the system
is helped to quickly converge to its state of minimum energy by a large damping
parameter $\alpha=1$. Starting from this magnetic configuration, the external
magnetic field is then applied strictly along \Fe{001} and the magnetization of
the system is monitored every picosecond during 20~ns by solving the \ac{LLG}
equation (\ref{eqn:llg}) with a timestep of 0.05~ps and a damping coefficient
$\alpha = 0.008$ \cite{tang2017, chen2018, khodadadi2020, zhang2020}.  The
magnetization dynamics in the time domain can be analyzed by a Fourier
transform to provide a power spectrum with a resolution of 0.05~GHz and the
phase \cite{mcmichael2005, baker2017}. The evolution of the system between
those two steps is small enough so as to consider its response as linear.

The red curve in figure~\ref{fig:fig4}a is the power spectrum for a uniform
system composed of clusters with identical magnetic properties coupled with
exchange constant $\tilde{J}=90$~meV. The spectrum shows a single peak at
6.45~GHz corresponding to the resonance phenomenon in the uniform film. This
peak position does not depend on the value of the exchange constant. For this
\textit{uniform} mode, all magnetic moments precess in phase around the
equilibrium position.  However, in a more realistic description, magnetic
properties are all slightly different from dot to dot.
This may be due to a dispersion of magnetoelastic or magnetocrystalline anisotropy
related to surfaces, interfaces or to the volume \cite{thomas2003}. These
inhomogeneities are taken into account by assuming a uniform random fluctuation
in the value of the perpendicular anisotropy energy which we model by adding a
small variation $\Delta K_{1,\perp}$ to the constant $K_{1,\perp}$.

Spatially averaged power spectra for $\Delta K_{1,\perp}V$=30~meV are shown in
figure~\ref{fig:fig4}a for some values of $\tilde{J}$. This means that the
perpendicular anisotropy energy change is randomly and uniformly distributed
over the [-30~meV; +30~meV] range. All spectra exhibit a main resonance peak
whose frequency decreases when $\tilde{J}$ is lowered. Introducing
inhomogeneities in the system leads to additional resonance peaks at higher
frequencies with larger relative intensities for weaker exchange constants.  It
is worth noticing that these features cannot be attributed to edge effects (the
so-called \textit{edge modes} \cite{mcmichael2005, guo2013})
since the system is periodic in \Vu{x} and \Vu{y} directions.  For
$\tilde{J}$=15~meV, the main resonance peak is at 4.60~GHz while the second
largest peak appears at 5.45~GHz. 

Looking at the phase values for each island
on the system is insightful since it reveals that all magnetic moments are
precessing in phase at the frequency of the main resonance peak while it is
clearly not uniform for the frequency corresponding to the secondary peak where
we can see two domains 40\degree\ out of phase with each other (maps of
figure~\ref{fig:fig4}b). The main resonance peak is the only one corresponding
to a uniform mode. All other smaller peaks at higher frequencies stemming from
the system inhomogeneity correspond to non-uniform modes. Those modes are
observed for any value of $\tilde{J}$ in an inhomogeneous system but the non
trivial analysis of their origin is beyond the scope of this work. We will thus
focus the rest of our study to the resonance peak of the uniform mode occurring
at a frequency \fres.
Let's see now how inhomogeneity of the islands' magnetic properties 
can affect the dynamic response of the system.

\subsubsection{Effect of inhomogeneous magnetic properties of islands}
\label{sec:fres_inhomogeneity}

\begin{figure}[htbp]
\includegraphics[width=86mm]{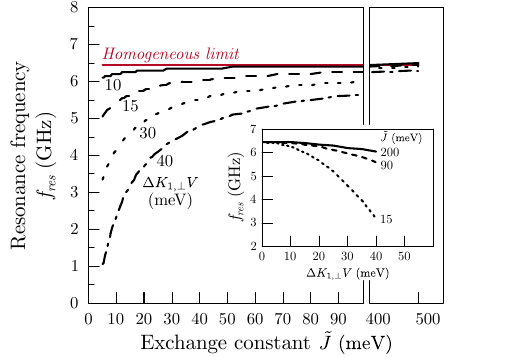}
\caption{\label{fig:fig5} 
    [Color online]
    Evolution of the resonant frequency $\fres$ versus $\tilde{J}$ for random
    fluctuations of the perpendicular anisotropy energy $\DKperp V$ between 10
    and 40~meV. The homogeneous limit corresponds to a film with $\DKperp V$ =
    0~meV. Inset: $\fres$ as a function of $\DKperp V$ for weakly interacting
    islands ($\tilde{J}$ = 15~meV), medium-coupled islands ($\tilde{J}$ =
    90~meV) or strongly-coupled islands ($\tilde{J}$ = 200~meV).
}
\end{figure}

The previous results highlight that introducing a reasonable
fluctuation in the value of the perpendicular anisotropy constant leads to
significant changes for \fres. We further examine this in figure~\ref{fig:fig5}
where the resonance frequency $\fres$ is plotted as a function of the exchange
constant $\tilde{J}$ (from 5 to 500~meV) for $\DKperp V = 10, 15, 30$ and
40~meV. 
The key observation is that $\fres$ drops when the system deviates from
being homogeneous.  These changes start to be significant ($>10\%$) for
$\DKperp V > 10$~meV (which is a rather small fluctuation of only $\sim 5\%$ to
the perpendicular anisotropy energy) and might be particularly noticeable for
$\DKperp V = 40$~meV and for weakly interacting clusters ($\tilde{J} = 5$~meV)
where $\fres$ is divided by $\sim$ 6.

The evolution of the frequency $\fres$ versus $\DKperp V$ for $\tilde{J} =
15$~meV (weak coupling), 90~meV (medium coupling) and 200~meV (strong coupling)
are reported in the inset of figure~\ref{fig:fig5}. Our simulations show that
there is a decrease in resonant frequency in the presence of lateral
inhomogeneity over short distances, compared to a film with uniform properties.
We find that we can retrieve the resonant frequency of such a system made up of
uniform islands for $\tilde{J} > 200$~meV.  Our results show that the impact of
these fluctuations will be all the greater if the clusters are only loosely
coupled to each other. The combination of inhomogeneity and weak exchange
coupling is indeed responsible for the reduction in resonant frequency compared
with a uniform system.

\subsubsection{Effect of islands' shape on the resonant frequency}
\label{sec:fres_shape}

A distribution of anisotropy constants due to some variability in magnetic
properties over the islands is not the sole source of inhomogeneity in such
systems.  The most obvious being the shape of the clusters that a close
inspection of experimental \ac{STM} images reveals that none of them are truly
circular (figure~\ref{fig:roughness}).  

To take into account the effect of the shape of the islands, we
attach a uniaxial anisotropy to each cluster whose direction is randomly and
uniformly distributed in the film plane.  We consider now that the islands are
cylinders with an elliptical base of major axis equal to the triangular lattice
parameter $a$ and of minor axis $b$. To save consistency with preceding
results, the magnetic volume is kept unchanged leading to an increase in island
height ($h_c$) by a factor $a/b$ (the aspect ratio). Computation of the
self-demagnetizing field tensor elements $N'_{xx}$, $N'_{yy}$ and $N'_{zz}$
respectively along the major, minor and \Vu{z} directions (for the major axis aligned
with \Vu{x}) are now all different \cite{beleggia2005} and two anisotropy
constants are now needed to describe the shape anisotropy of clusters: 
\textit{i)}  $\Kpshapeoop = \frac{1}{2}\mu_0 M_s^2 (2N'_{yy} + N'_{xx} - 1)$
             which is the updated version of the previously defined $\Kshapeoop$ and
\textit{ii)} $\Kpshapeip = \frac{1}{2}\mu_0 M_s^2(N'_{yy}-N'_{xx})$ which favors an
             in-plane magnetization along the major axis of the dot.
The magnetic anisotropy free energy density is slightly different from the previous
definition in equation~\ref{eq:EA1}:
\begin{equation}
    \label{eq:EA2}
    \begin{split}
        W_A = &-(K'_{1,\perp} + \Kpshapeoop)\cos^2\theta \\
              &- \Kpshapeip\sin^2\theta\cos^2(\phi - \phi_\textrm{ea}).
    \end{split}
\end{equation}
In the above equation, $\theta$ and $\phi$ are the polar and azimuthal angles
of the common spherical coordinates system.  $\phi_\textrm{ea}$ denotes the
angle between \Vu{x} and the major axis of the elliptical base corresponding to
the easy axis for magnetization.  As a little subtlety, introducing such shape
anisotropy will alter the $K_{1,\perp}$ term introduced in
equation~\ref{eq:EA1} since it is composed of a volume ($K_{{1,\perp}_V}$) and
a surface ($K_{{1,\perp}_S}$) parts: $K_{1,\perp} = K_{{1,\perp}_V} +
K_{{1,\perp}_S}/h_c$. As the latter is generally considered to be dominant,
$K_{1,\perp}$ is assumed to be inversely proportional to the island
height~\cite{bodker1994, bensch2002} which leads $K_{1,\perp}$ of
equation~\ref{eq:EA1} to be replaced by $K'_{1,\perp}\approx K_{1,\perp}
\frac{b}{a}$ in equation~\ref{eq:EA2}.

\begin{figure}[htbp]
\includegraphics[width=86mm]{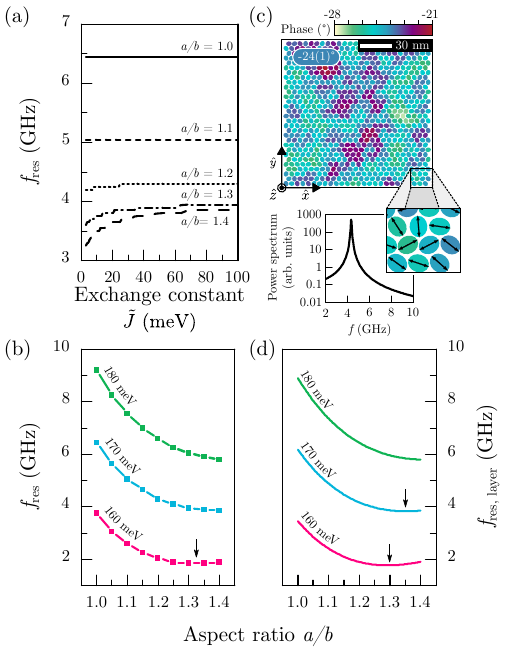}
\caption{\label{fig:fig6} 
    [Color online]
    a) Evolution of the resonant frequency $\fres$ as a function of the
    inter-islands effective exchange constant $\tilde{J}$ for increasing aspect
    ratio $a/b$ of the islands from 1.0 to 1.4.
    b) and d) Resonant frequency $\fres$ versus the aspect ratio $a/b$ as
    computed in the macrospin model with $\tilde{J}=90$~meV for $K_{1,\perp}
    V=160, 170, 180$~meV (b) and d) as given by equation~\ref{eq:layerfres}. 
    Small vertical arrows indicate curves minima.
    c) [top] Spatial map of the phase values for $K_{1,\perp}V=170$~meV,
       $\tilde{J}=90$~meV and $a/b=1.2$.
       Bridges are not drawn to avoid overloading the map.
       [bottom] The corresponding power spectrum as a function of the frequency 
       showing a single peak for this uniform mode.
}
\end{figure}

Figure~\ref{fig:fig6}a shows the evolution of the resonant frequency
$f_\mathrm{res}$ as a function of the exchange constant $\tilde{J}$ for aspect
ratio values in the [1.0, 1.4] range for $K_{1,\perp} V = 170$~meV. 
For $\tilde{J}$ values $> 90$~meV, the larger the aspect ratio, the lower the
resonant frequency. For example, $\fres$ is lowered by 40\% when the aspect
ratio goes from 1.0 to 1.4.  Our calculations also show that $\fres$ decreases
for lower values of the exchange coupling. 

In figure~\ref{fig:fig6}b, the resonant frequency $\fres$ is plotted as a
function of the aspect ratio for $K_{1,\perp}V$=160, 170 and 180~meV for
$\tilde{J}$=90~meV. Those values were injected in the simulation
and $K'_{1,\perp}$ values were deduced accordingly as previously discussed. We
observe a noticeable decrease of $\fres$ for growing aspect ratio for those
three $K_{1,\perp}V$ values. When the aspect ratio goes from 1.0 to 1.4,
$\fres$ is decreased by 37\% (respectively 40\% and 49\%) for
$K_{1,\perp}V$=180~meV (respectively 170 and 160~meV). Interestingly, the
variation is not only non linear, but also non monotonic since the curve for
$K_{1,\perp}V$=160~meV shows a minimum in $\fres$ for the 1.3 aspect ratio.
These results highlight how crucial it is to take into account shape
anisotropies of islands arising from the low scale roughness in ultrathin
magnetic films since those geometrical aspects may significantly alter the
resonant frequency of the system and consequently greatly impact the dynamical
properties of the materials.

To better understand the results of figure~\ref{fig:fig6}b, the spatial
distribution of phase values is drawn in figure~\ref{fig:fig6}c for the
simulation corresponding to an aspect ratio of 1.2 and for $\tilde{J}$=90~meV. The
bottom right corner of the map is zoomed in and black arrows parallel to major
axes are drawn to help in viewing how shape randomness is introduced in the
system.  The plot of the power spectrum versus frequency (bottom of the map in
figure~\ref{fig:fig6}c) exhibits only one peak without any other resonant modes
indicating that we are dealing with a uniform mode.  This is further confirmed
by the averaged phase value of -24\degree\ with a small standard deviation
of only 1\degree, showing that the phase is uniform for all islands.
Although for the lowest $\tilde{J}$ and the highest aspect ratio, peaks
corresponding to non-uniform modes begin to appear, they are barely
distinguishable.  Under these conditions, such a system may be regarded as a
unique macrospin precessing under an external magnetic field $H_0\Vu{z}$.
Let us recall the expression of the resonant frequency of an island having
demagnetizing field tensor coefficients $N'_{xx}$, $N'_{yy}$ and $N'_{zz}$ and an
anisotropy constant $K'_{1,\perp}$.
It reads:
\begin{equation}
    \label{eq:macrofres}
    \begin{split}
    f_\mathrm{res,\,island} = & \frac{\gamma}{2\pi M_s} \\
                              & \times \sqrt{2K'_{1,\perp} + \mu_0 M_s(H_0+M_s(N'_{yy}-N'_{zz}))} \\
                              & \times \sqrt{2K'_{1,\perp} + \mu_0 M_s(H_0+M_s(N'_{xx}-N'_{zz}))}
    \end{split}
\end{equation}
We suggest that the above expression~\ref{eq:macrofres} may be used to derive
an analytical form of the resonant frequency of the whole rough layer
$f_\mathrm{res,\,layer}$ that will be helpful to explain the shape of the
curves in figure~\ref{fig:fig6}b. Indeed, roughness will induce local magnetic
poles responsible of an in-plane demagnetizing field~\cite{schlomann1970,
zhao1999} leading tensor elements along \Vu{x} and \Vu{y} ($N^"_{xx}$ and
$N^"_{yy}$) to be non-zero in equation~\ref{eq:macrofres}. The layer is made up
of islands with an elliptical base whose major axes are randomly and uniformly
distributed in the plane, which cancels out any in-plane anisotropy, resulting
in $N^"_{xx}=N^"_{yy}$. To take into account magnetic dipole interactions
between islands, we computed the dipolar energy density $W_{\mathrm{ddi},x}$
($W_{\mathrm{ddi},z}$) corresponding to a saturated system in the \Vu{x}
(respectively \Vu{z}) direction.  Injecting this inter-islands dipole
interaction and the resulting isotropic nature of the in-plane magnetic
energy in the previous formula~\ref{eq:macrofres} leads to
\begin{equation}
    \label{eq:layerfres}
    \begin{split}
        f_\mathrm{res,\,layer} =  \frac{\gamma}{2\pi M_s} \Biggl[ & 2K'_{1,\perp} + \mu_0 M_s H_0 + \\
                                 & \mu_0 M_s^2 \left( \frac{1}{2} - \frac{3}{2} N'_{zz} - 2\frac{\Delta W_{\textrm{ddi}}}{\mu_0 M_s^2} \right) \Biggr],
    \end{split}
\end{equation}
where $\Delta W_\textrm{ddi} = W_{\textrm{ddi},z} - W_{\textrm{ddi},x}$.  Here,
the $N'_{zz}$ factor is still referring to a single isolated island.  Curves
of figure~\ref{fig:fig6}d were computed using the above
equation~\ref{eq:layerfres} for aspect ratio ranging from 1.0 to 1.4 and for
different values of $K_{1,\perp}V$.  We find a very good agreement with curves
obtained from the simulation (figure~\ref{fig:fig6}b) and we notably reproduce
the presence of a minimum for $\fres$ in the curve for $K_{1,\perp}V=160$~meV
(indicated by arrows in the figure).  This variation of $\fres$ is the result
of two opposing effects whose evolution with the aspect ratio does not
compensate for each other.  On the one hand, increasing the aspect ratio lowers
the area of the surface and interface of a dot, which in turn lowers the
contribution of the surface and interface anisotropies ($K'_{1,\perp}$ term in
equation~\ref{eq:layerfres}), leading to a decrease of the resonant frequency.
On the other hand, $N'_{zz}$ decreases for an increasing aspect ratio while
$\Delta W_\textrm{ddi}$ is still constant and tend to increase the frequency.

It is interesting to note that we have just operated a second scaling law
for the resonance frequency of a set of macrospins, which can now be described
on the basis of the resonance frequency of a single giant spin. The shape of the
islands and therefore the atomic-scale roughness of the film is contained in
the $K'_{1,\perp}$ and $N'_{zz}$ terms of equation~\ref{eq:layerfres}.

\subsubsection{Spin-wave propagation in inhomogeneous layers}
\label{sec:magnon}

Our calculations evidenced the impact of the low-scale roughness on
resonant phenomena in ultrathin layers.  We demonstrated that a combination of
a weak exchange interaction and fluctuating perpendicular anisotropy from dot
to dot will result in a significant drop of the resonant frequency of the
system along with the emergence and the enhancement of non uniform modes.  

As it was previously pointed out by several studies \cite{hicken1995,
prokop2009, bergman2010, devolder2016, yastremsky2019, brandt2021}, the
spin-wave stiffness constant (usually denoted $D$ in the literature) may
significantly be reduced for ultrathin ferromagnetic films compared to the
bulk. Interface effects \cite{brandt2021} or interdiffusion at interfaces
\cite{hicken1995, eyrich2014, sato2016} can cause this lowering of $D$.  

Based on our work, we can predict that the inhomogeneous nature of ultrathin
films may also alter the spin-wave stiffness constant relative to that expected
for the smooth counterpart. To be more quantitative, we have calculated the
backward volume spin-wave dispersion for in-plane magnetization in an
inhomogeneous iron waveguide 1~nm thick, 1000~nm long and 50~nm wide (data not
shown).  Dispersion curves were obtained by applying a magnetic excitation
varying as a cardinal sine in space and time following the method of reference
\cite{venkat2013}.  Our simulations showed that the spin-wave stiffness
constant is lowered by a factor of 6 when the effective exchange constant
$\tilde{J}$ drops from 90 to 15~meV. Such a linear behavior is consistent with
the atomistic picture where $D$ is proportional to the strength of the exchange
interaction \cite{vaz2008, sipr2019}.  

These results underline that the low scale roughness we consider in this study
will impact the spin-waves dispersion in wave guides of different shapes
\cite{venkat2013, mahmoud2020}.  It is then expected that experimental
measurements of spin-wave dispersion in waveguides made of ultrathin films
would reveal more about the strength of the coupling due to nanoconstrictions,
and would allow the determination of $\tilde{J}$.

\subsection{Effect of the temperature}
\label{sec:Temperature}

In this section, we will introduce the effect of the temperature in our
multiscale approach.  We will study how the magnetization and magnetic
anisotropy of an inhomogeneous iron ultrathin film evolve as a function of the
temperature.  We will show \textit{i)} how the strength of the inter-island
exchange coupling can impact the transition temperature from ferromagnetic to
paramagnetic states of the layer, and \textit{ii)} how the spin reorientation
temperature is affected by changes in the morphology of the film.
Magnetization dynamics will not be considered here since \ac{LLG} spin equation
of motion used in section \ref{sec:mag_dynamics} to simulate resonance
phenomena are no longer appropriate for the macrospin approach when temperature
rises close to the ferromagnetic/paramagnetic transition \cite{chubykalo2006}.
For such situations, the \ac{LLB} equation for finite temperatures is much more
pertinent \cite{garanin1997, chubykalo2006, kazantseva2008}, but is beyond the
scope of this article.  Atomistic calculations reported in this section are
performed using \ac{MC} methods with an adaptive Metropolis algorithm
\cite{alzate2019} and taking into account first- and second-neighbor exchange
interactions, with exchange constants $J_1$=27~meV and $J_2$=19~meV
respectively \cite{cardias2017}.

In this section, we will first study the evolution of the
magnetization and uniaxial anisotropy of an isolated island as a function of
the temperature within a classical spin atomistic approach. We will then
evaluate the exchange coupling between two islands as a function of the
temperature for various nanoconstriction shapes.  These obtained
temperature-dependent island magnetization, anisotropy and inter-island
coupling will then be used into our multiscale simulations of the magnetic
properties of an inhomogeneous ultrathin film.

\subsubsection{Temperature dependence of magnetization and anisotropy of an
isolated island} \label{sec:MsKu_vs_T} 

\begin{figure}[bthp]
\includegraphics[width=86mm]{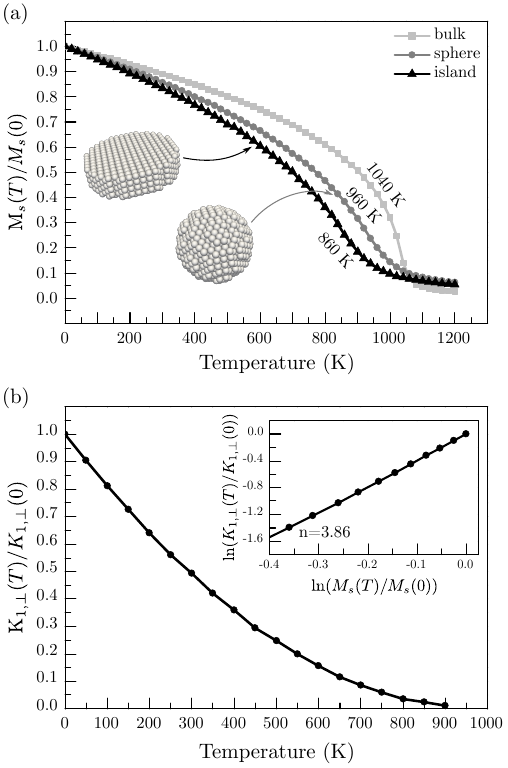}
\caption{\label{fig:figMsKu}
    a) Reduced magnetization versus temperature for bulk Fe (square symbol),
    for a single spherical island of 1606 atoms (full circle) and for a single
    disk-shaped island of 1604 atoms (triangle marker).  Transition
    temperatures from ferromagnetic order to paramagnetic state are also
    reported. b) Temperature dependence of the reduced anisotropy constant.
    Inset: log-log plot of the reduced anisotropy versus reduced magnetization
    (full circles) along with a linear fit to the data (black line). The slope
    $n$ is the exponent of the Callen and Callen model \cite{callen1966}.
}
\end{figure}

Let us first investigate the temperature dependence of the magnetization
($M_s(T)$) for an isolated iron island in a classical atomistic spin model. We
consider a disk-shaped \textit{bcc} Fe island 4.5~nm in diameter and 1.1~nm
high (8 atomic planes) containing $\sim$1600 atoms.  The evolution of $M_s(T)$
for such an island is reported in figure~\ref{fig:figMsKu}a along with the
magnetization curve of a bulk Fe lattice modeled as a cube comprising 16000
atoms with periodic boundary conditions.  Simulations are performed using the
\ac{MC} method with 10000 \ac{MCS} for relaxing the system followed by an
average over 100000 \ac{MCS}.  Magnetic anisotropy were not taken into account
in these calculations since $J_1$ and $J_2$ are much higher than the anisotropy
energies per atom.  We also excluded the \ac{DDI} contribution, since it plays
no role in our spin atomistic calculations.

We found the reduced magnetization curve $m(T)=M_s(T)/M_s(0)$ for the bulk
system to be very similar to that published by other groups using atomistic
simulations \cite{evans2015}. We extracted magnetic ordering temperatures from
the isothermal magnetic susceptibility as explained in \cite{landau2021}.  For
the bulk system, the transition temperature is 1040~K and is very close to the
experimental Curie temperature of 1044~K \cite{crangle1997}.  For the
disk-shaped island, magnetization is weaker than for the bulk, whatever the
temperature, and its transition temperature is $\sim$860~K.  The reduced
coordination number of atoms on the surface of the island compared to those in
its core is responsible for this lower transition temperature.  It is therefore
expected $m(T)$ curves to be sensitive to the island shape, as shown in
reference \cite{evans2006}.

To appreciate this shape-related sensitivity, Figure~\ref{fig:figMsKu}a also
shows the $m(T)$ curve for a spherical island containing $\sim$1600 atoms,
whose ratio of surface to volume atoms is lower than for the cylindrical
island.  Although the magnetization curve is close to that observed for a
cylindrical island, it decays less rapidly with $T$. We found $\sim$960~K for
the transition temperature, which is slightly higher than for the cylindrical
island.  In the following, we will only consider cylindrical islands.

We now focus on the temperature dependence of the island's anisotropy constant
$K_{1,\perp}$ which is assumed to be uniaxial in favor of an out-of-plane
magnetization due to symmetry breaking at
the surface, as described by L. Néel \cite{neel1954}. We will
disregard all other sources of anisotropy, such as magnetocrystalline volume or
magnetoelastic surface/volume anisotropies requiring far too much in-depth 
knowledge of the island shape, the deformation field within the islands as well as 
their temperature dependence.

We used \ac{CMC} simulations to determine the evolution of the reduced
anisotropy constant $K_{1,\perp}(T)/K_{1,\perp}(0)$ as a function of the
temperature \cite{sato2018, asselin2010}.  The \ac{CMC} algorithm is a slight
modification to the standard \ac{MC} technique. It allows to keep the mean
magnetization direction unchanged along a constrained direction (see
reference~\cite{asselin2010} for more details).  We implemented our own
\ac{CMC} method in the Spirit code to simulate the evolution of the reduced
anisotropy constant as a function of the temperature for a single cylindrical
dot.  Results are reported in figure~\ref{fig:figMsKu}b.  The inset also shows
this reduced anisotropy versus the reduced magnetization $m(T)=M_s(T)/M_s(0)$
in a log-log scale and clearly highlights the obtained linear variation as
predicted by the Callen and Callen model \cite{callen1966}.  Hence, for our
system, the reduced anisotropy can be described by a power law:
\begin{equation}
    \label{eqn:callen}
    \frac{\Koneperp(T)}{\Koneperp(0)} = \left[\frac{M_s(T)}{M_s(0)}\right]^n,
\end{equation}
with $n=3.86$. The value of the exponent $n$ is discussed in appendix~\ref{apx:callen_n}.
This analytical form will be used to parametrized the magnetocrystalline
anisotropy of multiscale calculations in section~\ref{sec:hetero_vsT}.

\subsubsection{Temperature dependence of the exchange coupling between islands}
\label{sec:coupling_vsT}

In this section, we will derive the temperature dependence of the exchange
coupling that exists between two cylindrical dots (as described in the previous
section~\ref{sec:MsKu_vs_T}) bonded by an atomic bridge.  We will use the
\ac{CMC} method to constrain the magnetization direction of the system while
allowing for thermal fluctuations, as was already done in
reference~\cite{evans2014} to calculate the strength of the exchange
interaction between two magnetic layers on either side of a magnetic
impurity-containing layer. The exchange interaction is derived from the angular
variation of the free energy, which is itself extracted from the thermodynamic
average of the magnetic torque \cite{asselin2010}:

\begin{equation}
    \label{eqn:torque}
    \VV{\tau} = \Big\langle \sum_i \hat{n}_i \times -\frac{\partial \mathcal{H}}{\partial \hat{n}_i} \Big\rangle
\end{equation}

The approach is similar to that described in section~\ref{sec:macrospins} at
T=0~K, but here we use the \ac{CMC} method and constrain the global
magnetization of one island along the $O\Vu{y}$ direction while that of the
second island is kept in the $\Vu{x}O\Vu{y}$ plane and defines an angle $\Phi$
with respect to the magnetization direction of the first island. No constraint
is imposed on spins inside the constriction. Magnetocrystalline anisotropy and
\ac{DDI} can safely be ignored in those calculations.  Convergence is achieved
after 10000 \ac{MCS}, and is followed by a series of 100000 additional \ac{MCS}
to compute the thermodynamic average of the torque.  To improve statistics,
torques are calculated using values from both islands. This procedure is
repeated every 10\degree\ for $\Phi$ varying from 0\degree\ to 180\degree.
Note that we had to use 10 times more \ac{MCS} for 180\degree\ to achieve the
necessary convergence and statistics.  In this configuration, the torque is
fully along $\Vu{z}$ which is why we will discuss this $\tau_z$ component in
the following.

\begin{figure}[tbp] 
    \includegraphics[width=86mm]{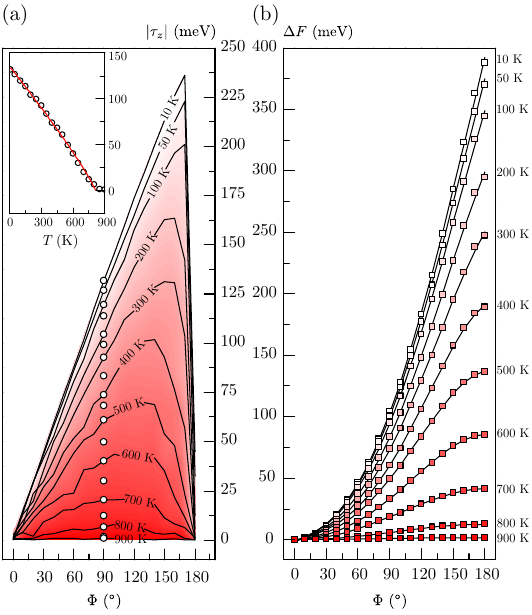}
    \caption{\label{fig:figTauz} 
        [Color online] a) Evolution of the magnetic torque as a function of the
        angle $\Phi$ between magnetizations of two islands of $\sim 1600$ atoms
        linked by a \textit{standard} atomic bridge of 15 atoms.  Contours (black lines) for
        temperatures ranging from 10~K to 900~K are also dsplayed and
        highlighted by the color gradient.  The 90\degree\ torque values marked
        by white circles are plotted in the inset as a function of the
        temperature. The red curve is the analytical fit suggested in
        section~\ref{sec:coupling_vsT}.  b) Free energy variation versus the
        angle $\Phi$ for temperatures between 10~K to 900~K as given
        by equation~\ref{eqn:deltaF}.
} \end{figure}

Absolute torque values $|\tau_z|$ versus the angle $\Phi$ for temperatures
ranging from 10~K to 850~K are shown in figure~\ref{fig:figTauz}a for a bridge
with dimensions $w=2$, $h_b=1$, $l=2$ in units of $a_0$. This bridge geometry
will be referred to as \textit{standard} in the following.  For $T<50$~K,
$|\tau_z|$ is linearly dependent on the angle $\Phi$ up to 170\degree\ and
quite abruptly drops between 170\degree\ and 180\degree.  At higher
temperatures, the $|\tau_z(\theta)|$ curve develops a bell shape and the
general trend is that torque magnitudes are lower when the temperature
increases, indicating a progressive decrease in the strength of the exchange
interaction between islands.

The inset of figure~\ref{fig:figTauz}a shows torque values at $\Phi=90\degree$
versus the temperature and reveals that $|\tau_z|$ decreases quasi linearly
with the temperature and falls to zero close to a temperature $\TCtwodots
\approx 812$~K. This temperature was determined by fitting the data points with
a function $f(T) = |\tau_z|_{90\degree}(0)(1-\frac{T}{\TCtwodots})^\beta$ where
$\beta$ is a fitting parameter (red curve in the inset of figure~\ref{fig:figTauz}a).

It is possible to relate the torque values $\tau_z(\Phi,T)$ to an effective
exchange constant $J_{12}(T)$ characteristic of the exchange interaction
between two islands having averaged magnetization directions \Vu{m_1} and
\Vu{m_2}.  For this, we need the free energy variation $\Delta F$ of the system
when the magnetization of the second island shifts by an angle $\Phi$ relative
to that of the first island \cite{asselin2010, evans2014, toga2016}:

\begin{equation}
    \label{eqn:deltaF}
    \Delta F(\Phi,T) = F(\Phi,T) - F(0,T) = - \int_0^\Phi \tau_z (\Phi',T) d\Phi'
\end{equation}

Neglecting magnetocrystalline anisotropies and \ac{DDI}, the conventional
Heisenberg expression also leads this free energy variation to be expressed as
the so-called \textit{bilinear term} \cite{slonczewski1993}:

\begin{equation}
    \label{eqn:bilinearterm}
    \Delta F(\Phi,T) = J_{12}(T)(1-\Vu{m_1}\Vu{m_2}) = J_{12}(T)(1-\cos(\Phi))
\end{equation}

Note that $J_{12}(0) = \tilde{J}$ used in sections~\ref{sec:macrospins} and
\ref{sec:mag_dynamics}.  Values of $\tilde{J}$ and \TCtwodots\ are reported in
table~\ref{tab:bridges} for all bridge geometries studied in this work.

From equations~\ref{eqn:deltaF} and \ref{eqn:bilinearterm}, it is then
straightforward to link the $z$-component of the torque to the $J_{12}$
parameter \cite{asselin2010, evans2014, toga2016}: 

\begin{equation}
    \label{eqn:tauzDF}
    \tau_z(\Phi,T) = - \frac{\partial \Delta F(\Phi,T)}{\partial \Phi} = -J_{12}(T)\sin(\Phi)
\end{equation}

According to this last expression~\ref{eqn:tauzDF}, and considering small
$\Phi$ angles, $\tau_z$ is expected to vary linearly with $\Phi$, as
$\sin(\Phi)\sim\Phi$ for small angles.  This is indeed what we observe for all
temperatures, and was also the conclusion we drew from our analysis at $T=0$~K
in section~\ref{sec:macrospins}.  However, this approach no longer holds when
$\Phi$ becomes significant.  Figure~\ref{fig:figTauz}a clearly shows that
$|\tau_z|(\Phi)$ cannot be described by a sine law.  Adding of a higher-order
term in the free energy (\textit{biquadratic term}), as proposed in reference
\cite{ellis2017}, does not improve this description either.

Therefore, to accurately account for the exchange interaction at all angles and
temperatures in multiscale calculations in section~\ref{sec:hetero_vsT}, we
will use an analytic derivable function (4\textsuperscript{th} order polynomial) to
closely match computed $\Delta F$ values at each temperature (full lines in
figure~\ref{fig:figTauz}).

\begin{table}[htbp]
    \begin{ruledtabular}
        \begin{tabular}{lrdldc}
            Bridge             & \multicolumn{3}{c}{Geometry}                    & \multicolumn{1}{c}{$J_{12}(0)$}  & \TCtwodots \\
                               & $h_b$   & \multicolumn{1}{c}{$w$}     & $l$     & \multicolumn{1}{c}{(or $\tilde{J}$)} &       \\
                               & [$a_0$] & \multicolumn{1}{c}{[$a_0$]} & [$a_0$] & \multicolumn{1}{c}{[meV]}        & [K]   \\
            \hline 
            \textit{standard}  & 1       & 2                           & 2       &  90     & 812 \\
            \textit{long}      & 1       & 2                           & 4       &  57     & 710 \\
            \textit{large}     & 2       & 2                           & 2       & 132     & 830 \\
        \end{tabular}
    \end{ruledtabular}
    \caption{
        Bridge geometries used in this study. Height ($h_b$), width ($w$) and
        length ($l$) are given in units of the iron lattice parameter $a_0$.
        The table also reports the effective exchange constant $J_{12}(T=0$~K$)=\tilde{J}$ 
        as well as the Curie temperature \TCtwodots\ for 2 cylindrical
        islands of $\sim 1600$ atoms connected by those atomic bridges.
    }
    \label{tab:bridges}
\end{table}

\subsubsection{Temperature dependent magnetization of inhomogeneous ultrathin film}
\label{sec:hetero_vsT}

In this section, we present the temperature dependence of the static magnetic
properties of an inhomogeneous ultrathin film in a multiscale approach.  We use
the \ac{MC} method and previous results of the inter-island coupling
$J_{12}(T)$, the island magnetization $M_s(T)$ as well as the anisotropy
$\Koneperp(T)$ from atomistic simulations are introduced.  More specifically,
we focus on the temperature dependence of the film's magnetization and magnetic
anisotropy. 

A number of previous studies have shown that film morphology influences the
temperature dependence of the magnetization of ferromagnetic films
\cite{schneider1990, ohresser1999, spangenberg2005, wenchin2006, bauer1997}.
For instance, a combined \ac{STM} and \ac{ACMOKE} study \cite{bauer1997} showed
that annealing a 4-monolayer Co film deposited at room temperature on a W(110)
surface led to an increase of its Curie temperature. This corresponds to the
transition from a layer made up of islands 4-5~nm in diameter to a smooth film
with large terraces $>$ 20-30~nm.  In addition, several studies have shown that
film morphology strongly influences \acp{SRT} in films. These transitions
correspond to a switch in the equilibrium direction of magnetization induced by
a variation in film temperature or thickness. A \ac{SRT} results from the
interplay of competing magnetic anisotropies favoring different easy axes. 

In this section, our aim is to examine the impact of roughness on static
magnetic properties of an ultrathin iron film $\sim$1.1~nm in nominal thickness.
As in section~\ref{sec:mag_dynamics} of this work, we will assume that the film
is composed of 780 equally spaced iron dots laid out on a triangular lattice of
parameter 4.5~nm, periodic along \Vu{x} and \Vu{y}.  Each island is modeled by
a macrospin whose saturation magnetization is $M_s(T)$, and the exchange 
interaction between
macrospins will be treated according to the exact approach discussed earlier
(section~\ref{sec:coupling_vsT}).  In these simulations, 50000~\ac{MCS} were
used to reach equilibrium followed by 50000~\ac{MCS} for averaging.

\begin{figure}[htbp] 
    \includegraphics[width=86mm]{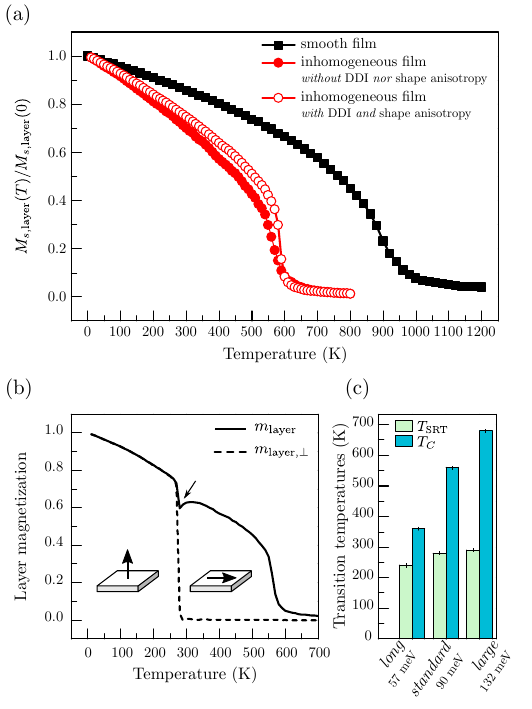}
    \caption{\label{fig:figTSRT} 
        [Color online] a) Reduced magnetization as a function of temperature
        for a smooth (homogeneous) thin film (black square) and for a rough
        (inhomogeneous) film composed of Fe islands with \textit{standard}
        bridges [in a) and b)] considering inter-island exchange only (full red
        circles) and adding \ac{DDI} and island shape anisotropies (open red
        circles). 
        b) Norm of the reduced magnetization versus the temperature (black
        line) and its $z$-component (dashed line) showing respectively the
        ferromagnetic-paramagnetic transition temperature \TClayer\ and the
        spin reorientation transition temperature \TSRT. Small drawings
        indicate the layer's magnetization direction.  c) Bar graph showing
        values of \TClayer\ and \TSRT\ for the three nanoconstriction
        geometries of this study (see table~\ref{tab:bridges}).  Error bars
        represents an uncertainty of $\pm 5$~K.  $J_{12}$ of bridges are also
        indicated in meV.
    }
\end{figure}

Figure~\ref{fig:figTSRT}a shows the reduced magnetization curve
$\Mslayer(T)/\Mslayer(0)$ calculated for an inhomogeneous ultrathin
layer (red full circles) compared with that obtained in an atomistic approach
for a uniform 8-atomic plane layer (black squares). In these simulations,
\ac{DDI}, shape anisotropies and magnetocrystalline anisotropies were not taken
into account. In the paramagnetic phase of the film, the magnetization of
macrospins remains non-zero as long as the temperature is below their order
temperature ($\sim$860~K, see figure~\ref{fig:figMsKu}). We observe that the
roughness leads to a drastic reduction in the film's transition temperature
\TClayer\ which drops from 910~K to 560~K. Note that this
transition temperature is still much lower than the transition temperature of
an isolated island.

For the sake of completeness, we also considered \ac{DDI} and shape anisotropy
of the islands ($\Kshapeoop V$) in our macrospin simulations (red open circles
in figure~\ref{fig:figTSRT}a). Although these additional terms
facilitate an in-plane magnetization, we can see in
figure~\ref{fig:figTSRT}a, that the magnetization curve is very similar to that
obtained by considering the exchange interaction only. One can note a slight
increase in the magnetic order temperature and a more pronounced curvature of
magnetization as a function of the temperature.  This increase in transition
temperature is to be expected, since in a two-dimensional triangular lattice,
\ac{DDI} enhance the ferromagnetic character of the lattice \cite{russier2001,
politi2002, politi2006}.

Our simulations highlight the significant impact of coupling strength on the
film's ferromagnetic to paramagnetic transition temperature, and allow a
quantitative characterization of this coupling. However, due to the lack of
available experimental data for measuring the inter-island exchange
interaction, a direct comparison with experiments is not possible.  Such a
determination of the inter-island coupling could be made by measuring the
dispersion of spin waves in waveguides made of inhomogeneous ultrathin films,
as we proposed earlier in this paper (section~\ref{sec:magnon}).

When the thickness increases or as the temperature is raised, ferromagnetic
thin films can undergo a \ac{SRT}, whereby the direction of magnetization
changes from out-of-plane (normal to the film surface) to in-plane (parallel to
the surface) (\cite{jensen2006} and references therein).  This transition
generally results from the competition between magnetocrystalline and
magnetoelastic anisotropies (favoring out-of-plane magnetization) and magnetic
dipolar interactions (promoting in-plane magnetization). 

To study this transition for inhomogeneous ultrathin films, we therefore add,
in our multiscale modeling, an additional surface anisotropy with anisotropy
constant $\Koneperp(T)$ and whose temperature dependence is the one we
calculated earlier (section~\ref{sec:hetero_vsT}). We took
$\Koneperp(0)V=230$~meV for islands 1.1~nm high and $\sim$4.5~nm in diameter.

Figure~\ref{fig:figTSRT}b shows the results of our simulations.  The normalized
magnetization of the layer $\mlayer(T)$ and its normalized $z$-component
$\mlayerperp(T)$ are plotted.  The overall layer magnetization
decreases as a function of temperature and is greatly reduced at 560~K,
corresponding to the ferromagnetic-paramagnetic transition.  This temperature
is close to the transition temperature of 580~K for the same system without
surface anisotropy. We note a singular point in the $\mlayer(T)$ curve
at $\sim$280~K (marked by an arrow on the figure) corresponding to the
\ac{SRT}. For $T<280$~K the magnetization is out-of-plane and for $T\ge 280$~K,
the perpendicular component of the layer magnetization drops to $\sim$0.  This
abrupt change in magnetization orientation is characteristic of a first-order
transition \cite{janke1998} and this discontinuity is consistent with
previously published work \cite{hucht1995}.

Let us now discuss how a change in the exchange coupling intensity may impact
both the ferromagnetic to paramagnetic transition temperature and the \ac{SRT}
temperature.  To this end, we carried out a series of simulations for the three
bridge geometries detailed in table~\ref{tab:bridges}.
Figure~\ref{fig:figTSRT}c shows the spin reorientation temperature
(\TSRT) and the ferromagnetic-paramagnetic transition temperature
(\TClayer) for those atomic bridges.  

\TClayer\ increases with increasing the exchange coupling. For the
\textit{long} bridge ($J_{12}(0)=57$~meV), this temperature is 360~K and rises
to 680~K for the \textit{large} bridge ($J_{12}(0)=132$~meV).  For a given
bridge geometry (\textit{standard} bridge), we find that \TClayer\ varies
between 420 and 630~K when changing the coupling strength by $\pm 50\%$, always
remaining below that of the smooth film.  Moreover, the transition temperature
does not vary linearly with the intensity of the exchange coupling (results not
shown).  This is different from what we observe in atomistic simulations, where
the transition temperature is proportional to the intensity of the exchange
interaction.  In our case, the inter-island exchange interaction is
temperature-dependent, which explains this difference.  These results show that
the ferromagnetic-paramagnetic transition temperature \TClayer\ is particularly
sensitive to the geometry of the bridges connecting the islands. 

The spin-reorientation temperature is also
impacted by changes in coupling intensity and increases as a function of the
exchange coupling magnitude, rising from 240~K for the \textit{long} atomic
bridge to 290~K for the \textit{large} one.

As already stated, the spin reorientation transition results from a competition
between magnetocrystalline anisotropy, which favors out-of-plane magnetization,
and the shape anisotropy of flattened disks, which favors in-plane
magnetization.  For its part, the long-range \ac{DDI} between macrospins also
promotes in-plane magnetization. These different contributions have different
temperature dependencies.  We found that the transition temperature depends
only little on the strength of the inter-island exchange coupling if we don't
take \ac{DDI}  into account,  meaning that the variation in \TSRT\ from
bridge to bridge, observed in figure~\ref{fig:figTSRT}c, is mainly related to
the variation in \ac{DDI} between the islands. Thermal fluctuations perturb the
orientation of individual magnetic moments, thus modifying the inter-island
\ac{DDI} when the coupling strength is altered.  For a given temperature,
fluctuations in the orientation of magnetic moments induced by thermal
agitation will be large if islands are loosely coupled, which in turn modifies
the inter-island \ac{DDI}.  We therefore show that the \ac{SRT} temperature
depends not only on the intrinsic properties of the islands, but also on the
nature of the inter-island contacts.

Finally, we investigated how the shape of an iron dot can change \TSRT.  For
this, we performed \ac{MC} simulations to calculate $\mlayer(T)$ for island
heights of 8, 9 and 10 atomic planes, while adjusting their radii so as to
maintain a constant number of atoms in each island. We considered a
\textit{standard} atomic bridge for these calculations.  As expected, the
ferromagnetic-paramagnetic transition temperatures of the films are virtually
identical for the three island heights ($\sim$560~K).  On the other hand, our
simulations show that spin-reorientation transition temperatures decrease from
280 to 240~K as the islands grow from 8 to 10 atomic planes (see
table~\ref{tab:TSRTvsBrigeHeight}).  These results demonstrate that the
transition temperature decreases with increasing roughness and that the
transition temperature will be higher the closer we get to a smooth film.

\begin{table}[htbp]
    \begin{ruledtabular}
        \begin{tabular}{lccc}
            Number of atomic planes & \multicolumn{1}{c}{radius} & \multicolumn{1}{c}{\TClayer} & \multicolumn{1}{c}{\TSRT} \\
                                    & \multicolumn{1}{c}{[nm]}  & \multicolumn{1}{c}{[K]} & \multicolumn{1}{c}{[K]} \\
            \hline
            8  & 2.28 & 560 & 280 \\
            9  & 2.18 & 560 & 270 \\
            10 & 2.03 & 560 & 240 \\
        \end{tabular}
    \end{ruledtabular}
    \caption{
        Ferromagnetic-paramagnetic transition temperatures \TClayer\ and spin reorientation temperature 
        \TSRT\ for $\sim 1600$ atom islands of 8 to 10 atomic planes connected with \textit{standard} atomic
        bridges.
    }
    \label{tab:TSRTvsBrigeHeight}
\end{table}

To summarize this section, we have presented the results of \ac{MC} simulations
designed to study the temperature dependence of the static magnetic properties
of an inhomogeneous ultrathin Fe layer with perpendicular magnetic anisotropy,
with a nominal thickness of 1.1~nm. We particularly focused on the evolution of
the spin-reorientation and ferromagnetic-paramagnetic transition temperatures
as a function of the layer's morphology.

A lot of published experimental studies \cite{schneider1990, ohresser1999,
spangenberg2005, wenchin2006, bauer1997} show that the Curie temperature
decreases for ferromagnetic ultrathin films with increasing roughness. Our
simulations well support these experimental observations and show that the
weaker the inter-island exchange coupling, the lower the Curie temperature.
Our model is consitent with experimental findings. This validates our approach.
We find that, for those systems, variations in island anisotropy energy and
\ac{DDI} barely have an impact on the value of the Curie temperature. In other
words, the Curie temperature of the inhomogeneous systems considered here
mainly depends on magnetic properties of the bridges linking the islands
together.

Concerning the spin reorientation transition, the literature is rich in
references pointing out the impact of the morphology on the \ac{SRT}
temperature \cite{schaller1999, kim2001, enders2003}.  Room temperature
\ac{MOKE} measurements on ultrathin Fe films on Ag(001) showed that increasing
the film's roughness leads to a decrease in the critical thickness below which
the layer shows perpendicular magnetization. This behavior is attributed to a
modification in magnetic dipole anisotropy and magnetocrystalline anisotropy
resulting from changes in surface morphology. Similar observations have been
reported for ultrathin cobalt films deposited on smooth and rough Pd surfaces
\cite{kim2001}.  Our simulations show that roughness impacts \ac{SRT} and well
support these experimental observations.

Although roughness can cover very different morphologies, our approach has
allowed us to quantify the key trends. In particular, we show that the greater
the aspect ratio of islands constituting the iron film, the lower the \ac{SRT}
temperature.  In our study, we highlight the role of inter-island exchange
coupling; in particular, we find that the transition temperature increases if
the intensity of inter-island coupling is increased. We find that the spin
reorientation transition temperature is higher the closer the system is to a
smooth surface.

\section{Discussion}
\label{sec:discussion}

We used a classical spin atomistic model and a multiscale approach to find out
the resonant modes as well as the order temperatures and \ac{SRT} temperatures
of a monocrystalline ultrathin film with perpendicular magnetic anisotropy made
up of nanometer-sized clusters of atoms. The shape of the clusters as well as
surface and interface magnetocrystalline anisotropies were taken into account
in our model, but most importantly we investigated how islands in intimate
contact through an atomic constriction or a discontinuous wetting layer are
interacting. This allowed us to give an origin and quantify the exchange-like
coupling between nearest-neighbor dots. Our approach is particularly well
suited to systems with complex and highly inhomogeneous morphologies where --
for example -- empty interstitial zones may be oddly distributed between
islands of magnetic matter. The approach can be used to describe
static/dynamical, in-plane/out-of-plane magnetic properties of such systems.

Given the particular importance of ultrathin magnetic layers for future
technological applications, there is no surprise that the scientific community
has devoted a great deal of effort to modeling them.  \Ac{FDM} \cite{zhu1988}
and \ac{FEM} \cite{lee2007} were formerly used, but they suffer from severe
shortcomings related to the space meshing required for any micromagnetic
calculation when it comes to describe highly disturbed surface morphologies at
low scale.  In \ac{FDM}, space has to be subdivided into cells of identical
sizes, generally limited by the material's exchange length. Here, however, the
size of these cells must be much smaller, not only to exactly reproduce the
irregular micro/nano structure of rough systems, but also to accurately resolve
the constrained magnetization within atomic constrictions, since its rapid
variation over few atoms is outside the hypothesis of continuous media for
exchange interaction \cite{andreas2014}. The computational power required is
therefore quite simply untractable.  FEM greatly improves this limitation by
the use of tetrahedral voxels whose size can be smartly adapted to local
variations in morphology.  But the technique will inevitably lead to the
over-discretization of islands with single-domain magnetization, unnecessarily
increasing the number of degrees of freedom making such simulations
particularly greedy in computation resources. However, innovative solutions
combining spin atomistic with \ac{FEM} \cite{andreas2014b} or with \ac{FDM}
\cite{de_lucia2016} have also been implemented and have made it possible to
accurately model specific spin textures like Bloch points or vortex cores.

It is also worth mentioning that micromagnetic solutions based on Voronoi
tessellation can be especially efficient in simulating the magnetic properties
of granular systems characterized by a significant structural complexity
\cite{peng2011, menarini2019, rannala2022}. Each grain is represented by a
unique Voronoi cell \cite{fidler2000, miura2005} and is further treated like a
macrospin.  This allows to reduce the number of unknowns while maintaining an
accurate description of the complex underlying microstructure of the granular
system. To describe inhomogeneity or polycristallinity in the material, grains
may be attributed some individual magnetic properties.  They can even be
separated by a non magnetic thin phase leading the inter-grain exchange
interaction to be driven by indirect mechanisms. In this description, grain to
grain exchange interaction is directly proportional to the contact area between
the grains \cite{peng2011, rannala2022}.

In our study, the exchange interaction between islands is not directly related
to the contact area between them. We showed that it is actually mediated by the
spins contained in atomic constrictions which eventually may be approximated by
a Heisenberg exchange interaction provided the angle between macrospin moments
does not exceed $\sim 45\degree$.  We also demonstrated that for higher
temperatures, the Heisenberg approach is no longer satisfactory and we used a
4\textsuperscript{th} order polynomial to fit the exchange free energy. This
allowed us to accurately take into account the inter-island exchange
interaction whatever the temperature and the angle between magnetization of
pair of islands The strength of the exchange interaction is closely related to
the geometry of the atomic constrictions with respect to the size of the
clusters. It should be noted that it also depends on their orientation with
respect to the crystal axes.  Indeed, due to a different number of interaction
pairs, the strength of the exchange interaction at 0~K is twice as large for a
constriction oriented at 30\degree\ with respect to the \Fe{100} direction as
for the same constriction along \Fe{100}, and up to three times as large in the
\Fe{110} direction.  In order to reduce the complexity, we have excluded this
orientation dependence from our study.

Our multiscale approach based on Monte Carlo calculations has allowed us to
model the temperature and thickness dependence of the static properties of
inhomogeneous ultrathin films. Magnetic order and spin reorientation
temperatures have been predicted in these systems, and we found a good
agreement between our simulations and trends derived from experimental
observations, validating our approach. Through our modeling we have been able
to unravel and identify the driving forces that control the temperature-driven
\ac{SRT} of inhomogeneous ultrathin films. We have limited ourselves to taking
into account magnetic dipolar interactions, surface magnetocrystalline
anisotropies and direct inter-island exchange coupling. 

However, the model could be improved by taking into account the precise shape
of the islands and surface/volume magnetocrystalline anisotropies (based on the
Néel anisotropy model for example \cite{neel1954, garanin2003, jamet2004,
yanes2007, morel2007, tournerie2008}), as well as the presence of island
deformations that will impact magnetic anisotropies through magnetoelastic
effects.  These improvements could obviously be added to our model without
creating any conceptual issue provided islands morphology is known in
sufficient detail.

Experimental observations show that clusters are tightly packed on the surface
and that each of them is surrounded by $\sim 6$ first neighbors. To reproduce
this in our study, we assumed those islands to be located on the nodes of a
triangular lattice, but the reality is not that regular and we could slightly
randomly shift their positions to improve our model. This will however
definitively eliminate the only periodicity that exists in the description of
these highly inhomogeneous systems, and will have an impact on the calculation
of the demagnetizing field, which remains the most difficult term in
equation~\ref{eqn:hamiltonian} to compute. The periodicity of lattices is
usually advantageously used to compute the demagnetizing field thanks to
powerful algorithms using \ac{FFT} techniques which is no
longer possible for a non-periodic system.  Other methods have to be used, such
as the \ac{NUFFT} \cite{menarini2019} or the \ac{FMM}~\cite{greengard1987, visscher2010}.

Finally, in our calculations, although the shape of an island is taken into
account through its shape anisotropy, its spatial extent is not, due to the
macrospin approximation which reduces an island to a point. The magnetostatic
field is thus computed within the point dipole approximation. We are currently
developing an improvement of our model which will be able to take into account
the true shape of the dots in the computation of the magnetostatic interactions
based on the fact that any shape can be magnetically equivalent to an ellipsoid
whose demagnetizing tensor can be analytically known \cite{beleggia2004,
beleggia2005}.

{\section{\label{sec:conclusions}Conclusions}} 

Exciting possibilities associated with spintronics for next-generation
electronics are closely related to magnetic properties of ultrathin films of
magnetic metals on metals, semiconductors and oxides as building blocks for many
devices. The theoretical effects of the inherent low-scale roughness of such
structures have long been challenging to assess, as neither standard
micromagnetism nor spin atomistic calculations can handle the large
computational resources needed to accurately describe these extensive
inhomogeneous systems.  

We have based our work on the modeling of the magnetic properties of a $\sim
1$~nm thick Fe(001) film with \acl{PMA} by parameterizing the morphology of the
system in the light of experimental data.  Using a classical spin atomistic
model, we have demonstrated that this highly disturbed surface morphology can
be represented by a compact ensemble of clusters bonded by atomic
constrictions, where constrained magnetic walls can couple clusters together. 

We have shown that such strongly inhomogeneous systems can be described by
exchange-coupled macrospins, whose strength can be precisely determined from
the shape of nanoconstrictions and from interatomic exchange constants.  

Reducing the clusters to macrospins allows for significant rescaling of the
system, making large-scale systems tractable. Through a multiscale approach and
solving the \ac{LLG} equation, we have simulated ferromagnetic
resonance spectra of inhomogeneous ultrathin films at 0~K. Our results show
that their magnetization dynamics is influenced by a distribution of uniaxial
magnetocrystalline anisotropy constants and of exchange constants, both of
which being key parameters characterizing low-scale roughness.

Additionally, we developed a multiscale approach based on \acl{MC} calculations
to model the temperature dependence of static properties of inhomogeneous
ultrathin films.  We then determined the ferromagnetic-paramagnetic and spin
reorientation transition temperatures for various morphology parameters,
allowing us to unravel and identify the driving forces controlling these
transition temperatures.  Our results are fully in line with experimental
observations, validating our approach.  More specifically, we have shown that
the spin reorientation transition temperature decreases as the roughness
increases, and that the Curie temperature is controlled by the strength of the
inter-island exchange interaction mediated by atoms in the nanoconstrictions.  

We believe that our approach demonstrates the possibility to account for the
morphology of ultrathin structures with significant roughness. Our approach
could be used to study the thermal stability and dynamic properties of devices
such as spin transfer torque magnetic random access memories integrating
ultrathin layers.
\\

\section*{\label{sec:acknowledgements}Acknowledgements}
The authors thank D. Sébilleau for sharing computing resources through a
project funded by the \ac{IEA}~2020 from the French National Centre for
Scientific Research (CNRS) as well as J.  Gardais and G. Raffy for their
technical support on the computing cluster.

\appendix
\section{Array of three dots}
\label{apx:three_dots}

\begin{figure}[htbp]
\includegraphics[width=86mm]{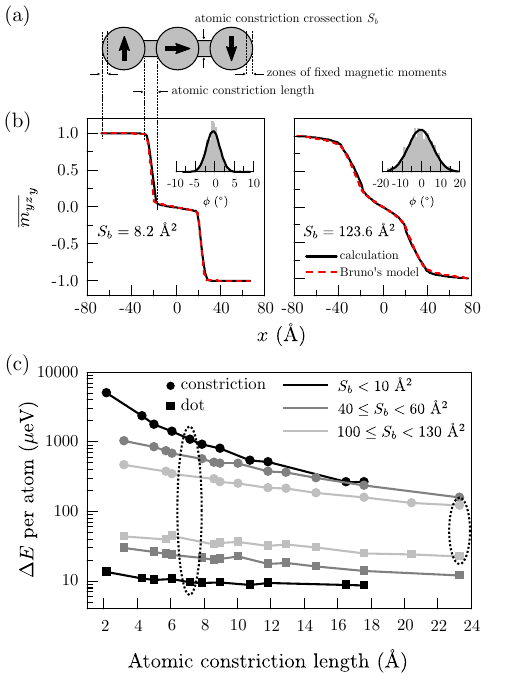}
\caption{\label{fig:bruno}
    [Color online]
    Magnetic configuration for an array of 3 dots linked by 2 atomic
    constrictions along $\Vu{x}$.
    a) Top view sketch of the 3 dots array where bold black arrows indicate the 
    magnetization direction in each cylinder.
    b) $y$-component of the magnetization averaged over the $yz$-plane as a
    function of $x$ coordinate with fixed magnetic moments at $x$ boundaries
    for a small constriction cross section [left plot] and a larger one [right
    plot]. Solid black curves are the calculation results and red dotted
    lines are the result of the Bruno model. Insets: histograms of the spins
    magnetization angles in the $xy$-plane (grey bars) along with a normal
    distribution fit to the data (black line) for the middle dot (-20~\AA $ < x
    < 20$~\AA).
    c) Mean exchange energy per atom for atoms in a dot [full square symbols]
    or in a constriction [full circles] as a function of the constriction
    length and for increasing cross sections. Points surrounded by a dotted
    line at $x\approx 7$~\AA (respectively 23~\AA) correspond to the geometry of
    figure~\ref{fig:bruno}a-left (respectively right).
}
\end{figure}

In this section, following the previous discussion in
section~\ref{sec:two_islands_magnetization}, we propose to investigate the
magnetic configurations for an array of three dots arranged along the
[100]$_{\rm{Fe}}$ direction and connected by nanoconstrictions
(figure~\ref{fig:bruno}a).  For such an array, magnetic moments of atoms lying
on $x$-boundaries over a distance of 2~\AA\ are fixed ($\phi = +90\degree$ on
the left hand side and $\phi = -90\degree$ on the right hand side), but spins
in the middle dot of the array can freely rotate.  The middle dot is connected
by two atomic constrictions (one to the left and one to the right).

Calculations similar to those in figure~\ref{fig:fig1}c were done for this
system where we also varied the geometry of both atomic bridges. We define the
shape of a bridge as a rectangular parallelepiped of length $l$ along the
\Vu{x} direction joining two dots, width $w$ along \Vu{y} and height $h_b$
parallel to \Vu{z}. The cross section area of an atomic constriction is
$S_b=w\times h_b$. The plane averaged magnetization $\overline{m_{yz}}_y$
component is plotted in the left (respectively right) panel of
figure~\ref{fig:bruno}b for atomic bridges 7.17~\AA\ (respectively 20.09~\AA) in
length and for cross sections of 8.2~\AA$^2$ (respectively 124.6~\AA$^2$).  We
see that for both cross sections, curves exhibit a staircase shape composed of
five quasi-linear sections, three of them corresponding to the islands with a
small slope as compared to the other two related to the atomic constrictions.
Clearly, the cross section area of the bridge changes how much rotation takes
place in both dots and constrictions.  This is especially visible on the right
plot of figure~\ref{fig:bruno}b. The distribution of the angle $\phi$ of the
magnetic moments in the middle dot is also reported in the insets of the figure
and confirms that spins are mostly parallel to each other but the larger the
bridge cross section area, the wider the standard deviation (1.7\degree\ and
6.3\degree\ for left and right panel respectively). 

These results demonstrate that the magnetization rotation is still mainly
absorbed by the atomic bridges.  Such constrained magnetic nanoconstrictions
were previously analytically investigated by Bruno \cite{bruno1999} for a
system with axial symmetry.  Following Bruno's model, and dropping any
magnetocrystalline anisotropy, the magnetization angle $\phi$ is given by:
\begin{equation}
    \label{eqn:brunomodel}
    \phi(x) = \pi \left[ 
                         \frac{1}{2} -
                         \frac{\int_{-\infty}^x S^{-1}(x')dx'}
                              {\int_{-\infty}^{+\infty} S^{-1}(x')dx'} 
                    \right],
\end{equation}
where $S$ is the analytical expression of the cross section of the system as a
function of $x$ split in 5 parts for the first dot ($S_0$), the first atomic bridge
($S_1$), the second dot ($S_2$), the second nanoconstriction ($S_3$) and the third
dot ($S_4$).
\begin{align}
    \begin{split}
    S_0 &= \frac{4}{3}\sqrt{r^2-x^2}, \; |x| \leq \delta \\
    S_1 &= wh + 2\left(\frac{2}{3}r - h\right)\sqrt{r^2-x^2}, \; \delta < |x| \leq r \\
    S_2 &= \frac{4}{3}r \sqrt{r^2-x^2},\; r < |x| \leq r+l \\
    S_3 &= S_1(x'), \; r+l < |x| \leq 2r+l-\delta \\
    S_4 &= S_0(x'), \; 2r+l < |x|
    \end{split}
    \label{eqn:S}
\end{align}

In formula \ref{eqn:brunomodel} and \ref{eqn:S},
$\delta = \sqrt{r^2-(w/2)^2}$ and $x' = 2r+l - |x|$.

Numerical integration of expression~\ref{eqn:brunomodel} yields the dashed line
curves reported in figure~\ref{fig:bruno}b. The agreement between atomistic
calculations and the model is stunning and support the conclusions of Bruno
that the magnetic configuration is only determined by geometric considerations:
ratio of cross section areas of the constriction and the dot as well as ratio
of constriction and dot lengths.  Magnetic properties of the material itself
won't affect the $\phi(x)$ profile of the magnetization.  To get an accurate
picture of exchange energies involved, we report in figure~\ref{fig:bruno}c the
energy difference per atom between the lowest energy configuration (all spins
parallel) and the most energetically expensive configuration (where
magnetization of end dots of the array are 180\degree\ apart).  Values for
atoms inside a constriction are given by full-circle symbols while those inside
a cluster are reported as full-square symbols. Results are given as a function
of the constriction length and cross section area. Points surrounded by the
dotted line at $\sim$7~\AA\ (respectively $\sim$20~\AA) correspond to the left
(respectively right) panel of figure~\ref{fig:bruno}b. We can see that for all
these geometries, the exchange energy per atom in a constriction is larger by
one or two orders of magnitude than for an atom inside an island. This energy
depends only little on the bridge length for atoms in a cluster but may be
multiplied by 3 when the constriction cross section area is multiplied
by~$\sim$10.

\section{Callen and Callen exponent}
\label{apx:callen_n}

The temperature dependence of the island's anisotropy constant $K_{1,\perp}$ is
assumed to be uniaxial in favor of an out-of-plane magnetization. According to
Callen and Callen \cite{callen1966}, the temperature dependence of the
anisotropy constant $K(T)$ for low-temperature ferromagnetic materials is
related to the magnetization $M_s(T)$ by a power law:
\begin{equation}
    \label{eqn:callen2}
    \frac{K(T)}{K(0)} = \left[\frac{M_s(T)}{M_s(0)}\right]^n,
\end{equation}
with $n=3$ for a uniaxial anisotropy and $n=10$ for a cubic anisotropy.
The previous relationship \ref{eqn:callen2} has already been successfully derived 
using \ac{CMC} simulations on continuous ferromagnetic layers with uniaxial
anisotropy, as well as on bulk magnetic systems with cubic anisotropy 
\cite{sato2018, asselin2010}.

We considered three distinct situations for the distribution of uniaxial anisotropy within 
the 8 atomic planes of the island numbered from 1 to 8 starting from the bottom surface:

\begin{enumerate}[i]
\item The anisotropy is distributed uniformly over atoms belonging to surface and interface only (planes 1 and 8)
\item The anisotropy is distributed uniformly over atoms belonging to planes 1, 2 and 7, 8
\item The anisotropy is distributed uniformly over all the volume of the island (planes 1 to 8)
\end{enumerate}

We computed the evolution of the reduced anisotropy
$K_{1,\perp}(T)/K_{1,\perp}(0)$ as a function of the temperature $T$ for the
three aforementioned cases by using the \ac{CMC} method. We found the
variation of the reduced anisotropy versus the reduced magnetization
$m(T)=M_s(T)/M_s(0)$ in a log-log scale to be perfectly linear for all three
cases, as predicted by the Callen and Callen theory.  We extracted the exponent
$n$ by fitting the data with equation~\ref{eqn:callen2} and the results are
given in table~\ref{tab:nvalues}.

Values for cases \textit{ii)} and \textit{iii)} are in good agreement with the
theoretical value of 3 for uniaxial anisotropy.
In case \textit{i)} $n=3.86$. Although slightly above the expected value, we find $n\sim
3$ when considering only magnetization of the surface planes in
equation~\ref{eqn:callen2} \cite{ibrahim2022}.  

\begin{table}[htbp]
    \begin{ruledtabular}
        \begin{tabular}{rlr}
            \multicolumn{2}{c}{anisotropy distribution} & $n$ \\
            \hline
            \textit{i)}   & planes 1 and 8     & 3.86 \\
            \textit{ii)}  & planes 1,2,7 and 8 & 3.25 \\
            \textit{iii)} & planes 1 to 8      & 2.81 \\
        \end{tabular}
    \end{ruledtabular}
    \caption{Exponent values $n$ of the Callen and Callen model (equation~\ref{eqn:callen2})
    for three different distributions of the anisotropy inside an island.}
    \label{tab:nvalues}
\end{table}

\bibliography{biblio}

\end{document}